\documentclass[preprint,showpacs,preprintnumbers,amsmath,amssymb]{revtex4}
\usepackage[dvips]{graphicx}

\begin{document}

\title{Charge Exchange Spin-Dipole  
 Excitations of $^{90}$Zr and $^{208}$Pb and Neutron Matter
 Equation of State }   
\author{ H.  Sagawa~$^{\rm a}$,  Satoshi Yoshida~$^{\rm b}$, 
Xian-Rong Zhou$^{\rm c}$, 
K. Yako~$^{\rm d}$, and H. Sakai~$^{\rm d}$}
\address{$^{\rm a}$~Center for Mathematical Sciences, the University of Aizu, 
Aizu-Wakamatsu, Fukushima 965-8580, Japan\\
$^{\rm b}$~Science Research Center, Hosei University 
2-17-1 Fujimi, Chiyoda, Tokyo 102-8160, Japan \\
$^{\rm c}$ Department of Physics and Institute of Theoretical Physics and Astrophysics, Xiamen University,
Xiamen 361005, P. R. China\\
$^{\rm d}$ Department of Physics, University of Tokyo, 
 Bunkyo, Tokyo 113-0033,
 Japan\\
}
\date{\today}

\begin{abstract}
 Charge exchange spin-dipole (SD) excitations 
of $^{90}$Zr and $^{208}$Pb 
are studied by using a Skyrme 
 Hartree-Fock(HF) + Random Phase approximation (RPA).
The calculated spin-dipole strength distributions 
 are compared with experimental
data obtained by $^{90}$Zr (p,n)  $^{90}$Nb and 
 $^{90}$Zr (n,p) $^{90}$ Nb reactions. 
 The model-independent SD sum rule values of various Skyrme
 interactions are studied in comparison with the experimental values in order 
 to determine the neutron skin thickness of 
  $^{90}$Zr.  
The pressure of 
 the neutron matter equation of state (EOS) and the nuclear 
 matter symmetry energy are discussed in terms of
the neutron skin thickness and peak energies of SD strength distributions.

\end{abstract}
\maketitle

\newpage

\section{Introduction}

The relationship between the neutron matter equation of state (EOS) and 
the neutron skin thickness has been studied extensively by using 
the Skyrme Hartree-Fock (HF) model, a relativistic mean field (RMF) model
\cite{Brown,Furn,Yoshi04}. 
 The neutron matter EOS is essential for studying the 
properties of neutron stars, e.g., their size
 \cite{Prakash}. It is also known that isovector nuclear matter
 properties, including the symmetry energy, correlate strongly
with the neutron skin thickness in heavy nuclei 
 \cite{Furn,Dani,Yoshi06}.

Elastic electron scattering has provided accurate data on the
charge distributions of nuclei. Several experimental attempts have been
made to measure neutron distributions, for example, 
 by proton elastic 
scattering \cite{pscatt1,pscatt2,Hoffman,Starodubsky}
 and by inelastic alpha scattering to 
giant dipole resonance excitations \cite{GDR}.  
However, empirical results of 
neutron skin thickness obtained by proton scattering are 
controversial and do not agree with each other even within experimental
error. The accuracy of empirical data on neutron distributions from giant 
resonance experiments is also rather poor, insufficient to extract accurate
information on the neutron matter EOS. One promising tool for 
studying neutron distributions is the parity violation electron 
scattering experiment \cite{escatt}.  
  Unfortunately, no data on parity violation electron 
scattering experiments is available so far.

 The model-independent sum rule strength of charge exchange 
SD excitation is directly related to
 information on the neutron skin thickness\cite{Gaarde}. 
Recently, SD excitations were studied in $^{90}$Zr by the charge 
 exchange reactions 
 $^{90}$Zr(p,n)$^{90}$Nb \cite{Wakasa} and  $^{90}$Zr(n,p)$^{90}$Y 
 \cite{Yako}, and the model-independent sum rule strengths for the
SD excitations were extracted in Ref. \cite{Yako06}  by using multipole
decomposition (MD) analysis  \cite{Ichi}.
 The charge exchange reactions 
($^3$He,$t$) on Sn isotopes
were also studied to extract the neutron skin thickness 
 \cite{SD-Pb}.
However, one needs the counter experiment ($t, ^3$He) or (n,p)
on Sn isotopes 
in order to extract
the model-independent sum rule value from experimental data. This counter 
experiment is missing in the case of Sn isotopes.

It is known that the SD strength has almost the same amount of contributions
to neutrino reactions as that of the Gamow-Teller strength \cite{Suzuki}.
The Pb target is considered to be the most promising candidate for detecting 
the heavy-flavor neutrinos from the supernovae. Thus, it
is quite important to study the SD strength in the Pb target for a  
 precise evaluation of the cross-sections of charge-induced neutrino 
reactions.

In this paper, we study the SD excitations  
and the neutron skin thickness by using the 
HF and HF+ random phase approximation (RPA) with Skyrme interactions.  
As a theoretical model, the HF+RPA model
  has been extensively applied to giant resonances 
in a broad region of mass table\cite{Ber75,HSZ97}. 
  The same model was used for the study of  
   spin-dependent charge exchange excitations\cite{SG2,Colo,SSG,HS2000}. 
It was shown that the model successfully predicts GT and SD states
 in $^{48}$Sc and $^{90}$Nb\cite{Colo,SSG}.     
First, we calculate the SD states in nuclei with mass A=90 and 208
by using the charge exchange HF + RPA 
model with various Skyrme interactions. We will compare  
calculated results of SD strength distributions with
 empirical data obtained by 
  charge exchange (p,n) and (n,p) reactions on $^{90}$Zr.
  The sum rule values are also compared with the empirical values in 
Section 2.
  Next, the correlations between the neutron matter EOS and 
the SD sum rules are studied in the Skyrme HF model.
We will discuss the neutron matter EOS by using the experimental
SD data and other empirical information on the neutron skin.
This paper is organized as follows.
In Section 2, the SD strength 
  of the HF+RPA  calculations 
  is  presented 
for both the $t_-$ and $t_+$ isospin channels on $^{90}$Zr and $^{208}$Pb.
  The calculated results are 
compared with experimental results of $^{90}$Zr(p,n)$^{90}$Nb
and $^{90}$Zr(n,p)$^{90}$Y reactions.
We study the correlations between the sum rules of  SD strength
and the pressure of neutron matter EOS in Section 3. 
A summary is given in section 4.

\section{HF+RPA calculations of SD strength}
  The operators for SD transitions are defined as 
\begin{eqnarray}
\hat{S}_{\pm} &=& \sum_{im\mu} t_{\pm}^{i}\sigma_{m}^{i} r_{i}Y_{1}^{\mu}
(\hat{r}_{i})
\label{eq:eq1}
\end{eqnarray} 
 with the isospin operators $t_{3} =  t_{z}, 
 t_{\pm} =  (t_{x}\pm it_{y})$. 
The model-independent sum rule for the
  $\lambda -$pole SD operator $\hat{S}^{\lambda}_{\pm } =
 \sum_{i} t_{\pm}^{i}$ 
$r_{i}[\sigma \times Y_{1}(\hat{r}_{i})]^{\lambda}$  can be 
obtained as 
\begin{eqnarray}
 S^{\lambda}_{-}- S^{\lambda}_{+}&=&
 \sum _{i \in all} \mid \langle i\mid\ \hat{S}^{\lambda}_{-} \mid0\rangle
 \mid ^2 -
\sum _{i \in all} \mid \langle i\mid\ \hat{S}^{\lambda}_{+} \mid0\rangle
 \mid ^2      \nonumber \\
&=& \langle 0\mid\ [\hat{S}^{\lambda}_{-}, \hat{S}^{\lambda}_{+}] \mid0\rangle
 = \frac{(2\lambda+1)}{4\pi}(N\langle r^2\rangle _n -Z\langle r^2\rangle _p).
 \label{eq:sum_sd_a}
\end{eqnarray}
 The sum rule for the spin-dipole operator (\ref{eq:eq1})
then becomes
\begin{eqnarray}
 S_{-}- S_{+}=\sum_{\lambda}(S^{\lambda}_{-}- S^{\lambda}_{+})=
\frac{9}{4\pi}(N\langle r^2\rangle _n -Z\langle r^2\rangle _p).
 \label{eq:sum_sd}
\end{eqnarray}
It should be noted that 
the sum rule (\ref{eq:sum_sd}) is directly related to the difference between
the mean square radius of neutrons and protons 
with the weight of neutron and 
proton numbers.  
We adopt four Skyrme interactions, namely, SIII, SGII, SkI3 and SLy4, 
for the HF+RPA
  calculations.  The Landau parameters and nuclear matter properties 
 of these interactions 
are  shown in Table \ref{tab:landau}. For the spin-isospin excitations,
the value $G_0'$ plays the important role of determining the collective
properties of the excitation \cite{SG2}.
The RPA equation is solved using the basis expanded by the harmonic
oscillator wave functions up to the maximum major quantum number of $N_{max}$
=10 for $^{90}$Zr and $N_{max}$=12 for $^{208}$Pb. 
 The HF calculations are performed without the spin-gradient terms 
(J$^2$ terms) since the adopted Skyrme interactions have been fitted
 without them \cite{Bei,SG2}, but  the RPA calculations incorporate
  the spin-gradient  terms.   The two-body spin-orbit and
two-body Coulomb interactions are neglected in the RPA calculations. 
We also performed the continuum HF+RPA calculations with one of the
interactions and found essentially the same strength 
distributions as in the present calculations except 
the width due to the coupling to the continuum
 \cite{HSZ97}.
The calculated results are smoothed out 
by using a weighting function, $\rho $:
\begin{equation}
  \frac{dB(SD)_{ave}}{dE_x}=\int \frac{dB(SD)}{dE_x '}\rho(E_x '-E_x)dE_x '
\label{eq:ave}
\end{equation}
where the weighting function is defined as 
\begin{equation}
\rho(E_x '-E_x)=\frac{1}{\pi}\frac{\Delta/2}{(E_x '-E_x)^2+(\Delta/2)^2}
\label{eq:weight}
\end{equation}
taking the width parameter $\Delta$.  
In the present calculations with the discrete basis, the SD strength is given by
\begin{equation}
 \frac{dB(SD)}{dE_x '}=\sum_i B(SD;E_i)\delta(E_i-E_x ').  \nonumber
\end{equation}

\begin{table}[htp]
\footnotesize\rm
\caption{\label{tab:landau}
 Landau parameters,  effective mass $m^*$ and symmetry energy J 
of Skyrme interactions
}
\begin{tabular}{l|c|c|c|c}  
\hline
 & SIII & SGII & SkI3 &SLy4 \\ \hline
$F_0$    &  0.309  & -0.235 & -0.318   & -0.273\\
$F_0'$  & 0.862   & 0.733 & 0.653  & 0.818 \\
$G_0 $   &  0.052 & 0.014 & 0.569  & 1.120 \\ 
$G_0'$ &  0.457 & 0.509 &   0.203 & -0.138 \\ \hline
$F_1$    & -0.709  & -0.646 & -1.269  &-0.926 \\
$F_1'$  & 0.490  & 0.521  &-0.843   & -0.399\\
$G_1 $   & 0.490  & 0.612 & 1.33  & 0.279 \\ 
$G_1'$ & 0.490  & 0.432   & 0.65  & 1.047\\ \hline
$m^*/m$& 0.76 &0.78 & 0.58  & 0.69 \\
J(MeV) & 28.1 & 26.9 & 34.8  & 32.3 \\ \hline
\end{tabular}
\end{table}

\subsection{Charge exchange SD excitations of $^{90}$Zr}

\begin{figure}[htp]
\vspace{-2cm}
\includegraphics[width=3.0in,clip]{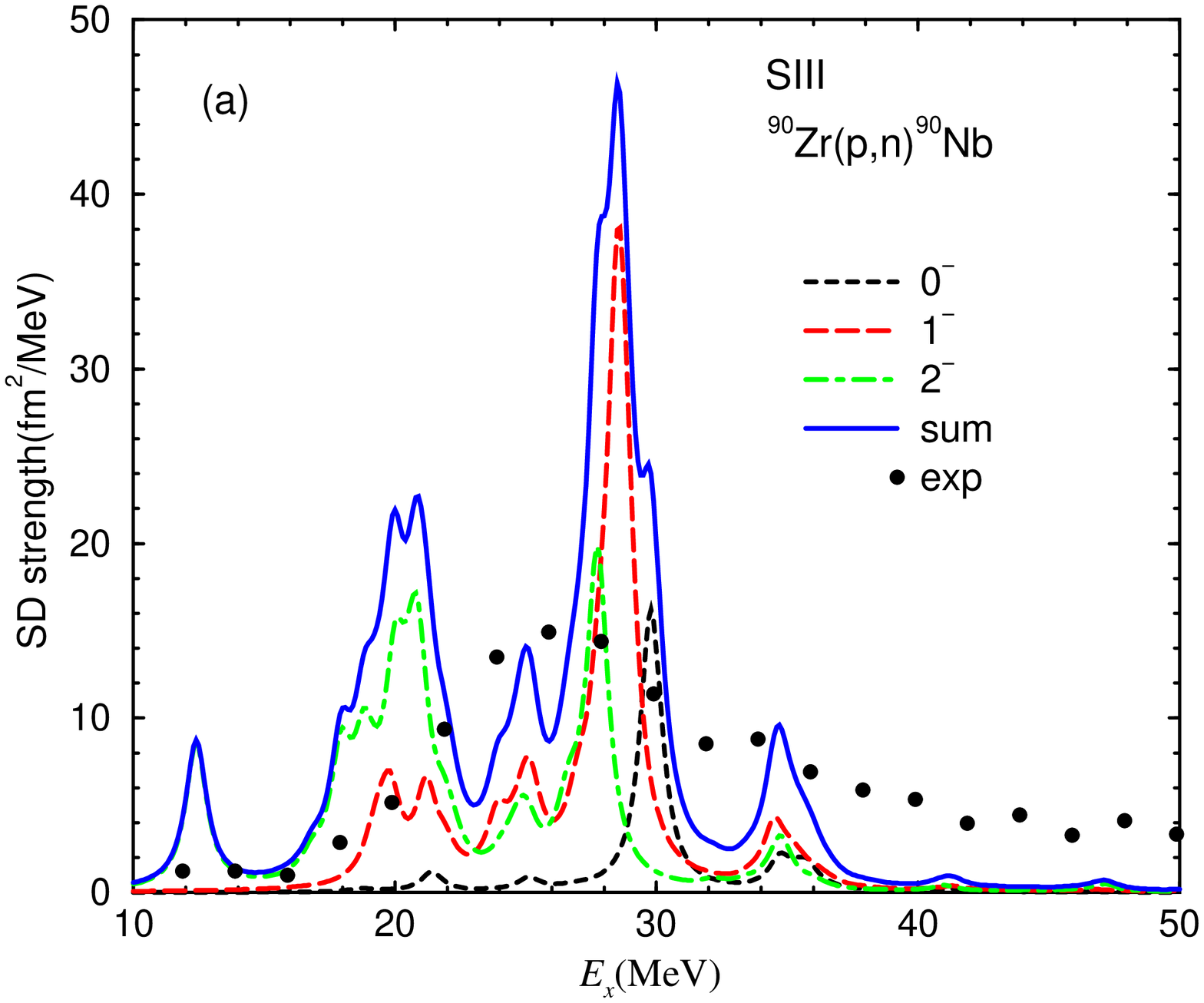}
\vspace{-2.0cm}
\includegraphics[width=3.0in,clip]{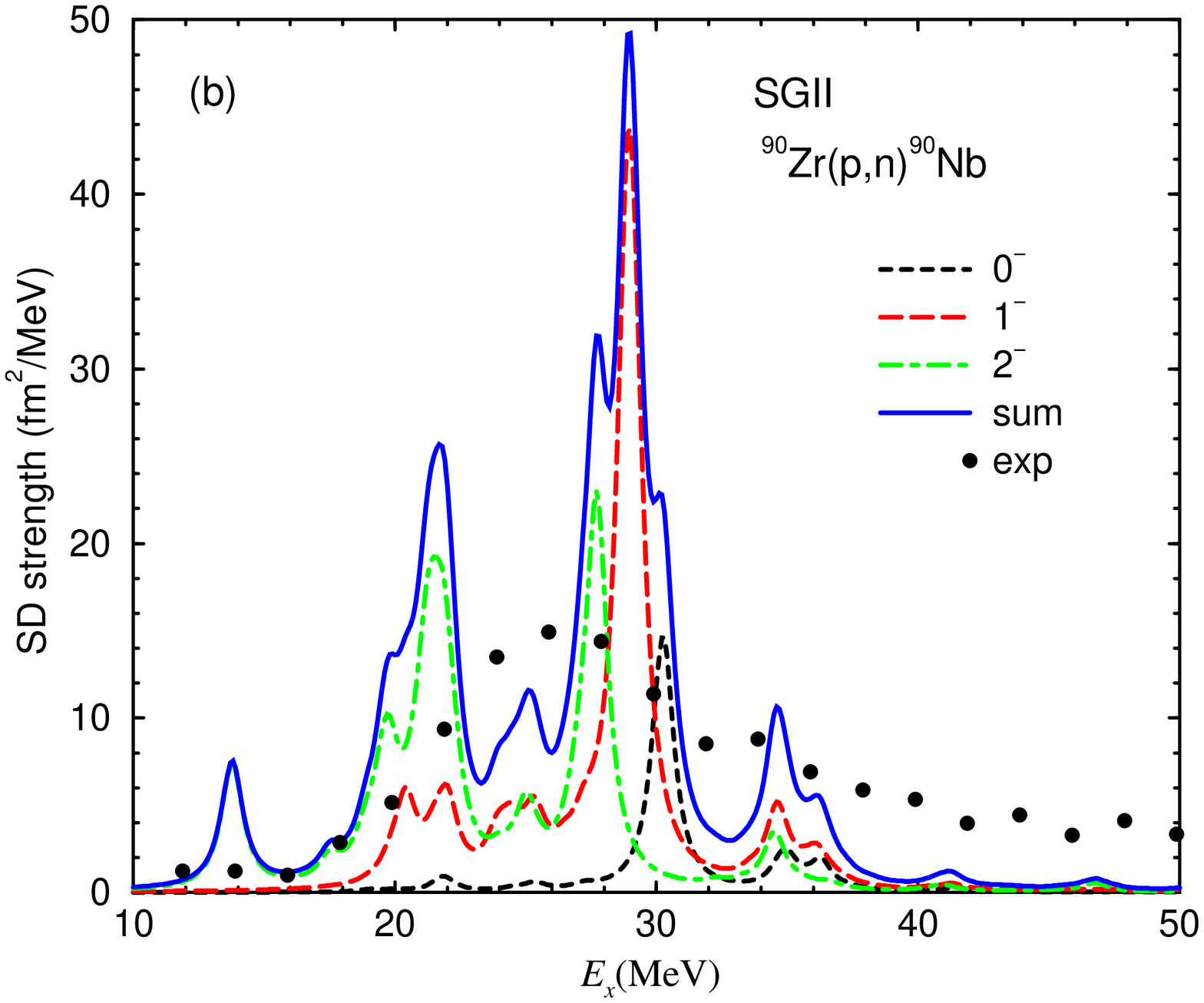}
\includegraphics[width=3.0in,clip]{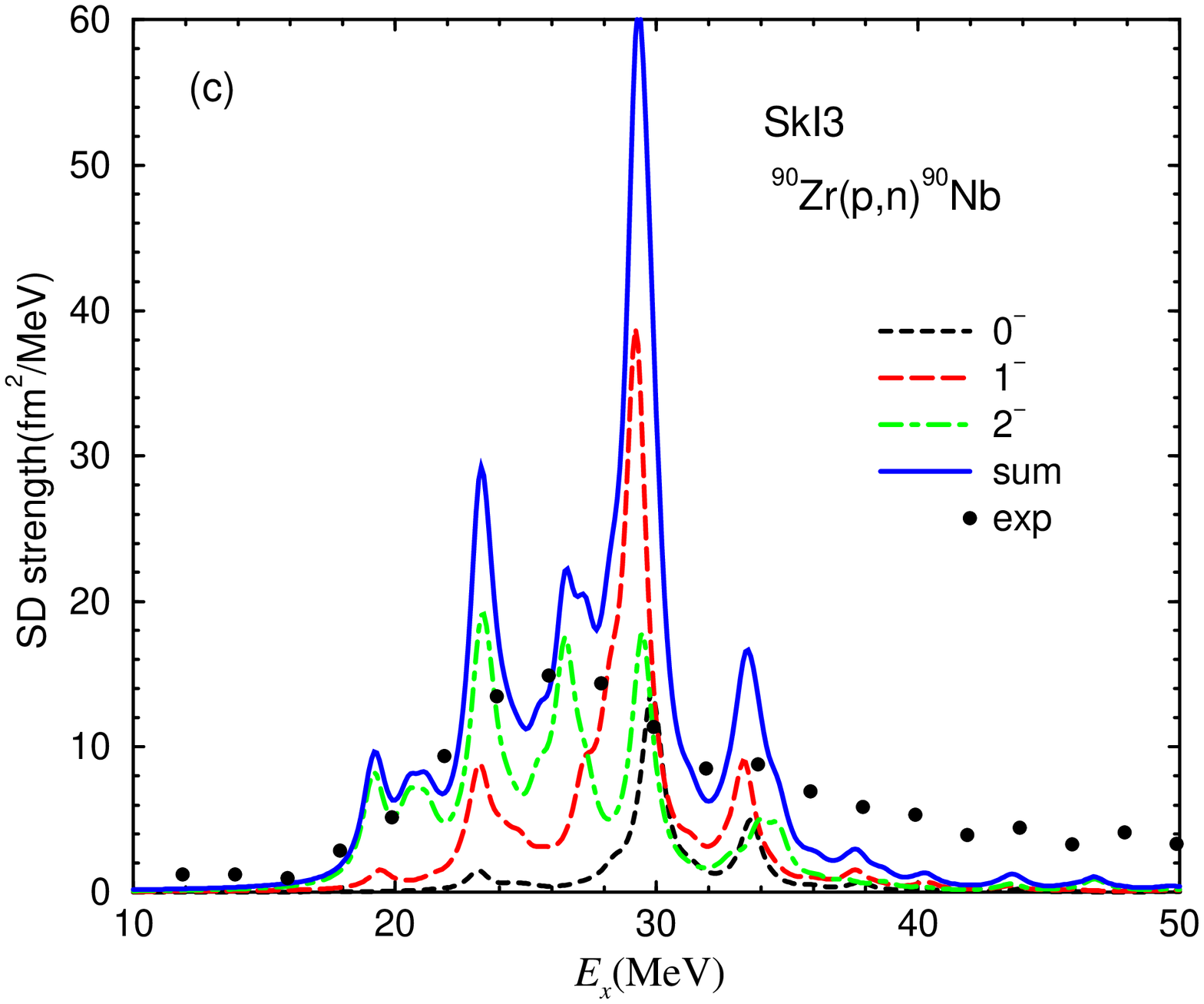}
\includegraphics[width=3.0in,clip]{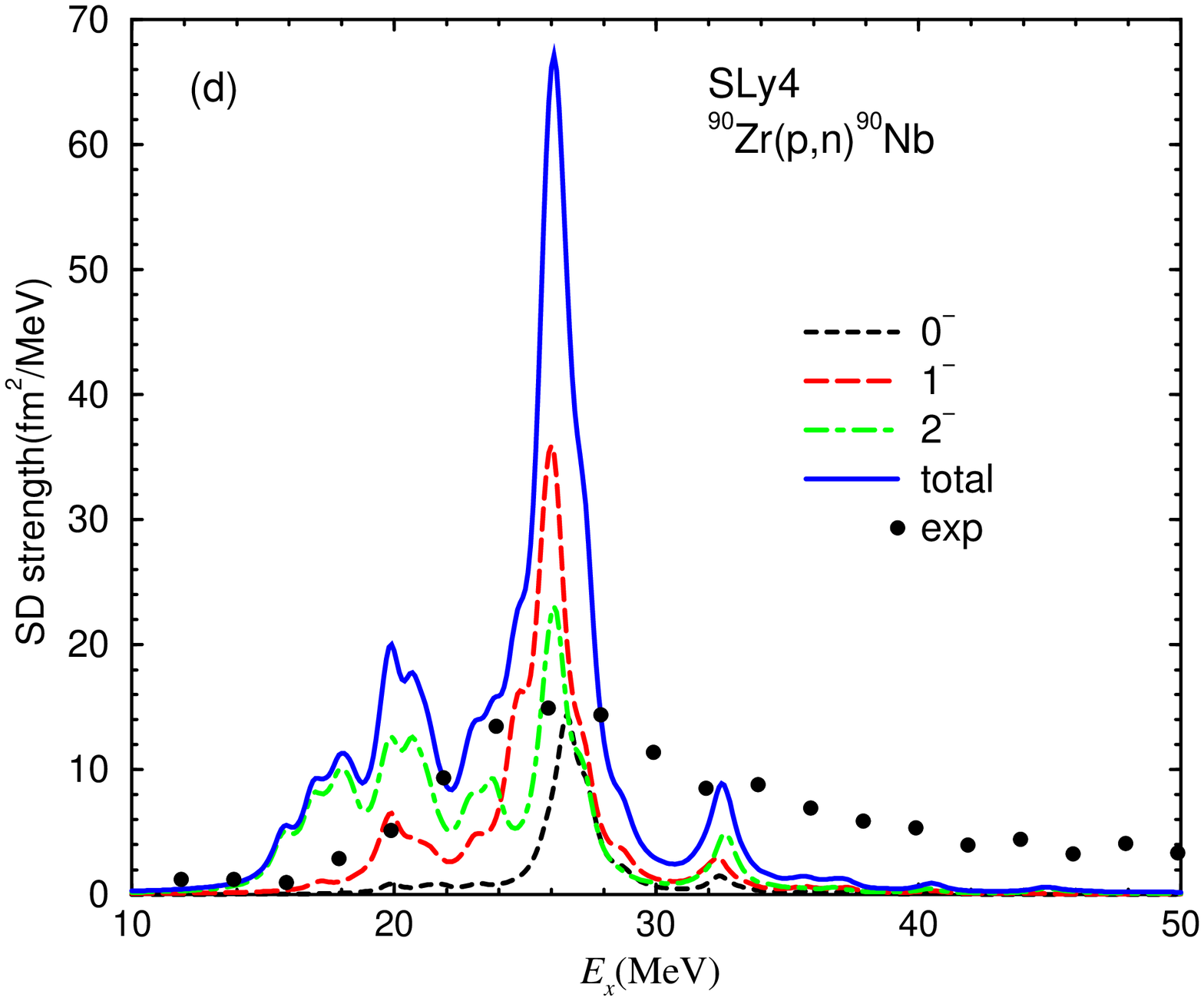}
\caption{\label{fig:zr90_sdm}(Color online) 
Charge exchange SD strengths for 
  the operators $\hat{S}^{\lambda}_{-} =
 \sum_{i} t_{-}^{i}$ 
$r_{i}[\sigma \times Y_{1}(\hat{r}_{i})]^{\lambda}$ 
calculated by the HF+RPA model with 
the Skyrme interactions (a) SIII, (b) SGII, (c) SkI3 and (d) SLy4.
The excitation energy is referred to the ground state of the
parent nucleus $^{90}$Zr.  
The dotted, dashed and long$-$dashed lines show the SD strengths of
$\lambda=0^-, 1^-$ and $2^-$ , respectively, while the solid curve shows 
the sum of three multipoles.
The SD strength is averaged by the weighting function (\ref{eq:weight}) 
  with the width 
$\Delta $=1MeV.  The experimental data shown by the 
 black dots are taken from ref. \cite{Yako06}.}
 \end{figure}

\begin{figure}[htp]
\vspace{-2cm}
\includegraphics[width=3in,clip]{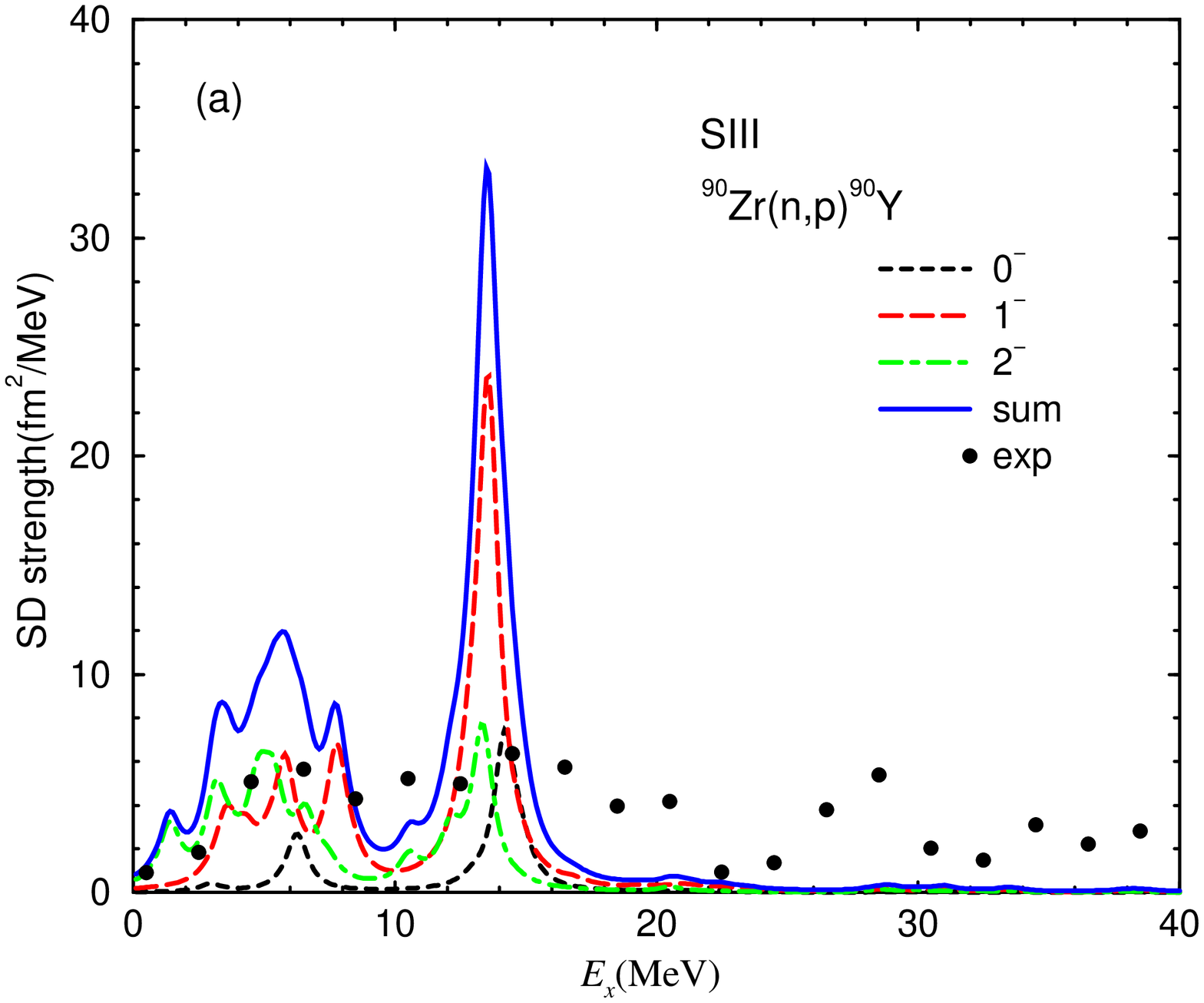}
\vspace{-2cm}
\includegraphics[width=3in,clip]{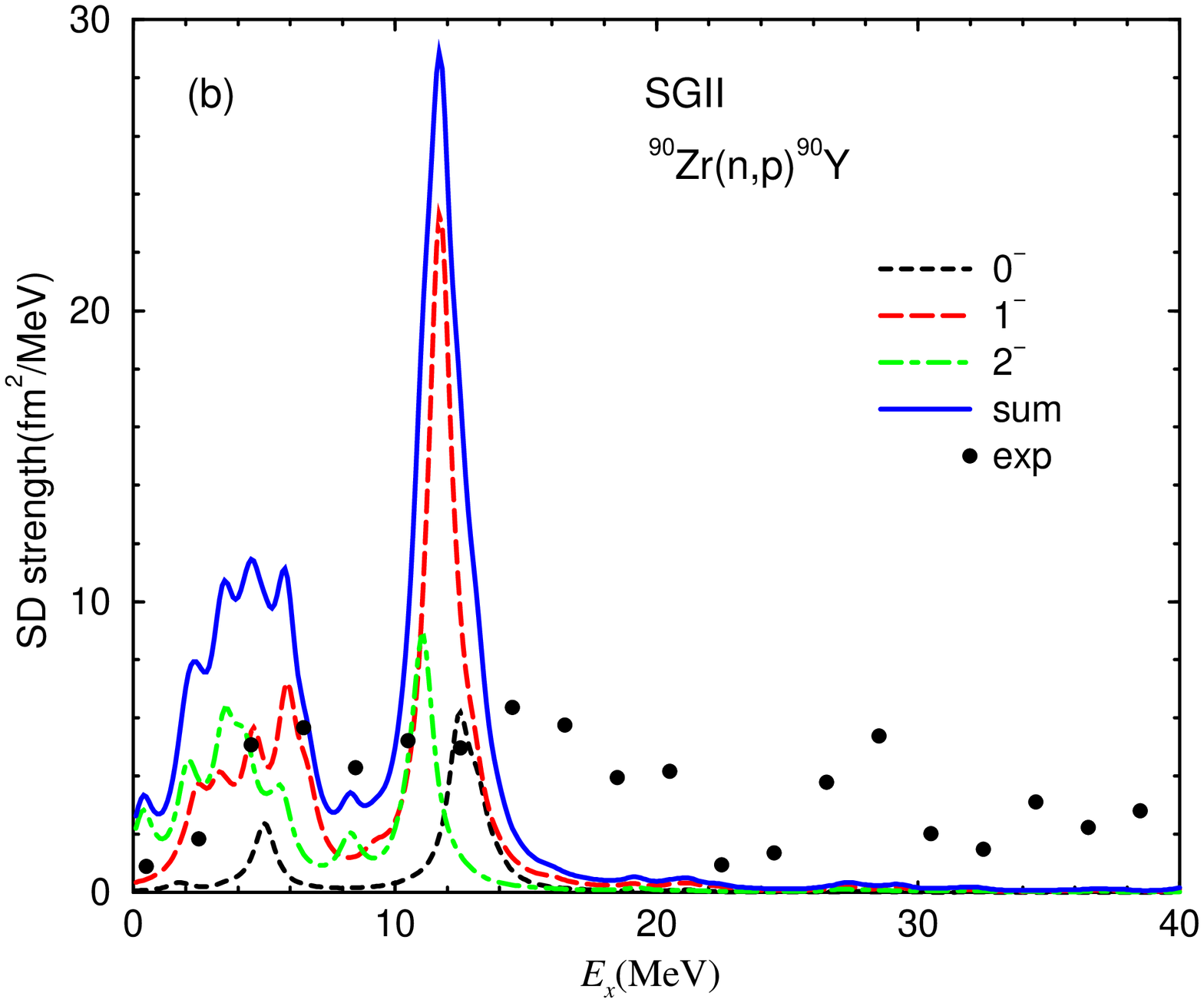}
\includegraphics[width=3in,clip]{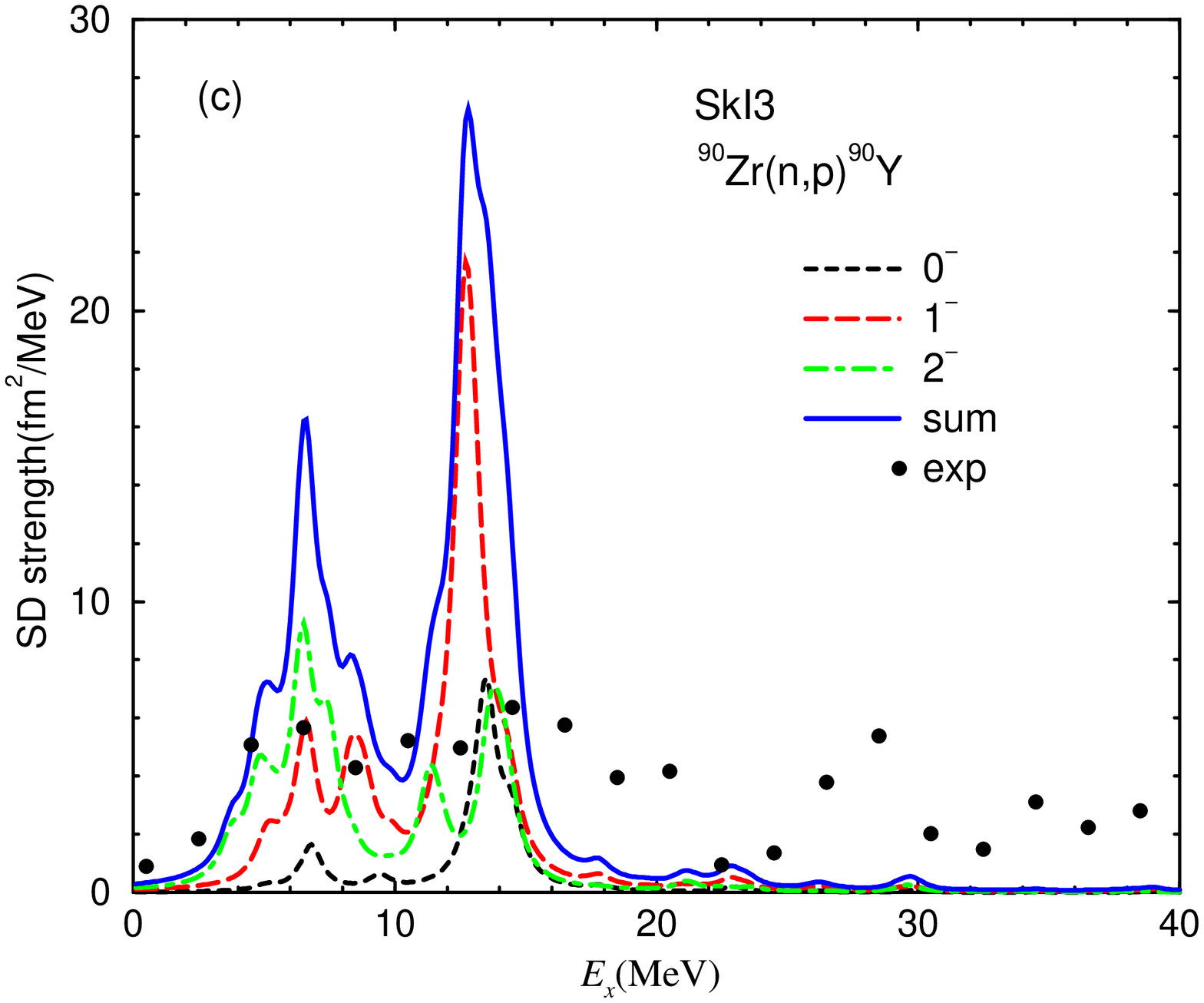}
\includegraphics[width=3in,clip]{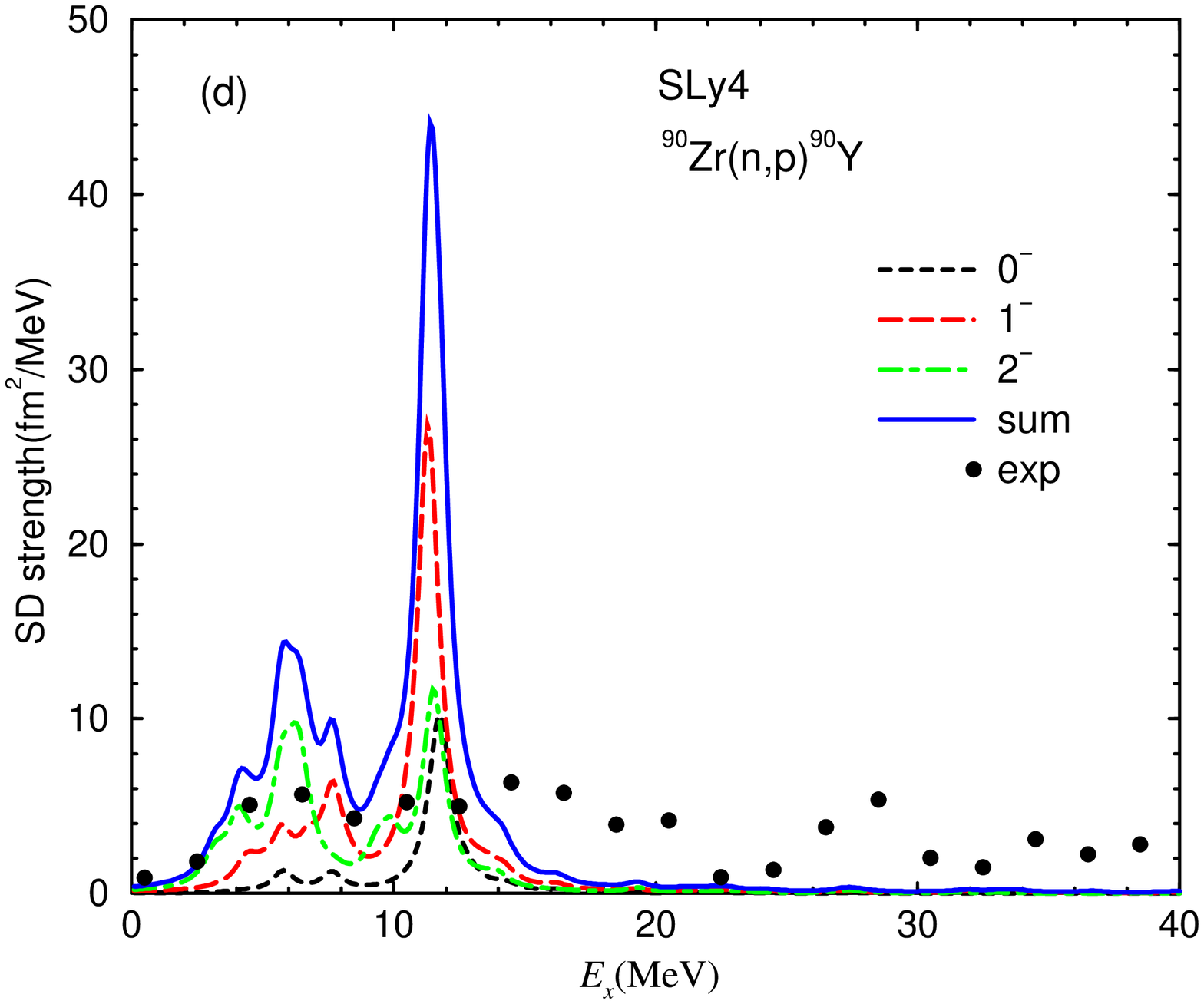}
\caption{\label{fig:zr90_sdp}(Color online)
Charge exchange SD strengths for the operators $\hat{S}^{\lambda}_{+} =
 \sum_{i} t_{+}^{i}$ 
$r_{i}[\sigma \times Y_{1}(\hat{r}_{i})]^{\lambda}$ 
calculated by the HF+RPA model using
the Skyrme interactions (a) SIII, (b) SGII,  (c) SkI3 and (d)SLy4.
The excitation energy is referred to the ground state of the
parent nucleus $^{90}$Zr. 
The SD strength is averaged by the weighting function (\ref{eq:weight})
  with a width of $\Delta $=1MeV.
The experimental data shown by the   
 black dots are taken from ref. \cite{Yako06}. 
 See the captions to Fig. \ref{fig:zr90_sdm} for details.}
 \end{figure}

The HF calculations are performed by using four Skyrme 
interactions in Table \ref{tab:landau}. The proton, charge and neutron radii of  $^{90}$ Zr are
  listed in Table \ref{tab:hf-zr}
   together with the sum rule values $\Delta S=S_--S_+$ calculated through the 
analytic equation (\ref{eq:sum_sd}).  By using the same HF wave functions, 
the charge exchange RPA calculations give the
 SD strengths in $^{90}$Nb and $^{90}$Y excited by
the  $t_{\pm} r\sigma Y_{1}(\hat{r})$ operators from the parent nucleus
  $^{90}$Zr,  as  shown in
 Figs. \ref{fig:zr90_sdm} and \ref{fig:zr90_sdp}.  
The experimentally obtained distributions of the SD strengths
are also plotted in Figs. \ref{fig:zr90_sdm} and \ref{fig:zr90_sdp}.
The experimental SD strength distributions for the $t_-$ and
the $t_+$ channels were obtained from the $^{90}$Zr(p,n)$^{90}$Nb
and the $^{90}$Zr(n,p)$^{90}$Y data, respectively, by performing
MD analysis \cite{Yako06}. A comprehensive description of the MD
analysis can be found in ref. \cite{Ichi}.

\begin{table}[htp]
\caption{\label{tab:hf-zr}
 Proton, neutron and charge radii of $^{90}$Zr.
 The charge radius is obtained by folding the proton finite size.
 The sum rule values $\Delta S=S_--S_+$ of spin-dipole excitations are 
  calculated 
 by Eq. (\ref{eq:sum_sd}) with the HF neutron and proton mean square radii.  
Experimental data on the charge radius are taken from ref. \cite{Vries}.
The experimental values $r_n-r_p$ are taken from \cite{pscatt1,Yako06}. 
 The radii are given in 
units of fm, while the SD sum rules are given in units of fm$^2$.}
\begin{tabular}{l|c|c|c|c|c}   
 \hline
 & SIII & SGII & SkI3 &SLy4 &exp \\ \hline
$r_p$ &  4.257 & 4.198 & 4.174 & 4.225  & ----\\
$r_c$  & 4.321 & 4.263  & 4.240 & 4.290 &4.258$\pm$0.008\\
$r_n$ &  4.312  & 4.253 & 4.280 & 4.287& ----  \\ \hline
$r_n-r_p$ & 0.055 & 0.055 & 0.106 & 0.064 &0.09$\pm$0.07 \cite{pscatt1}, 
 \,\,\, 0.07$\pm$0.04 \cite{Yako06}   \\ \hline
$\Delta S$ & 146.7 & 142.9 & 156.9 & 146.9 &  \\
\hline
\end{tabular}
\end{table}

\begin{table}[htp]
\caption{\label{tab:Epeak}
  Peak energies and the average energies 
 of charge exchange SD excitations in the A=90 nuclei
obtained by the self-consistent HF+RPA calculations: $t_-$ in $^{90}$Nb and
 $t_+$ in  $^{90}$Y. The average energy is calculated 
by the ratio of EWSR to NEWSR: \=E(MeV)=$m_1/m_0$. See the text 
for details.}
\footnotesize\rm
\begin{tabular}{l|c|c|c|c} 
\hline
  &\multicolumn{2}{c|}{$t_-$} &\multicolumn{2}{c}{$t_+$} \\ \hline
   & $E_{peak}$(MeV) & \=E(MeV)&  $E_{peak}$(MeV) & \=E(MeV) \\ \hline
 SIII & 28.5 & 25.7 & 13.5 & 10.9 \\
 SGII & 27.7 & 26.7 & 11.7 & 9.47 \\
 SkI3 & 29.3 & 28.2 & 12.8 &  11.6 \\
  SLy4 & 26.1 & 24.9 & 11.4 & 10.5 \\ \hline
\end{tabular}
\end{table}


In general,
the $t_-$ SD strength distributions 
 for 0$^{-}$ and 1$^{-}$ states in $^{90}$Nb
 are concentrated in one
state at $E_{x}\sim$30MeV, 
having a large portion of the non$-$energy weighted sum 
rule (NEWSR) strength, while those for the 2$^{-}$ states
are separated into two dominant peaks, as shown in Fig. \ref{fig:zr90_sdm}.
 The   0$^{-}$ peak appears at $E_x\sim$30MeV, having 73\%, 65\%, and 58\%
 of 
the NEWSR value for the SIII, SGII and  SkI3 
   interactions, respectively.
The  calculated results for 1$^{-}$ states show a peak at $E_x\sim$29MeV 
having  50\%, 59\%  and   48\% of the NEWSR value
  for the SIII, SGII and SkI3   
  interactions, respectively. The three
 results in Figs. \ref{fig:zr90_sdm}
 (a), (b) and (c) show the 0$^{-}$ peak
 at a very similar excitation energy, 
 while the values of NEWSR    are somewhat different. 
 The same is true 
for the 1 $^{-}$ peak in the three results.
For the SLy4 interaction 
  in Fig. \ref{fig:zr90_sdm}(d), the 0$^{-}$ and 1$^{-}$ peaks 
appear at about 3 MeV lower than the other three results, having
76\%  and 68\% of the NEWSR, respectively.
This is due to the negative value of  the Landau parameter $G_0'$ in
  SLy4 for the spin-isospin channel.
The dominant configurations of the collective  0$^{-}$ and 1$^{-}$ states
 are the $(\pi1h_{9/2}\nu1g_{9/2}^{-1})$ and  $(\pi1g_{7/2}\nu1f_{7/2}^{-1})$ 
configurations.
For the 2$^-$ excitations, the number of p-h configurations is larger than 
those of 0$^{-}$ and 1$^{-}$ and therefore 
 the strength is fragmented in a wider energy range 
compared with 0$^{-}$ and  1$^{-}$ excitations.
There is a small low-lying peak with $J^{\pi}=2^-$ 
 at $E_{x}$ = 12.4 (14.1) MeV with
 10.0 (9.0)\% of the  NEWSR value in the case of the SIII (SGII) interaction.
This state is mainly due to the 
 $\pi1g_{9/2}\nu1f_{5/2}^{-1}$ configuration.
The major strengths are found in the two peaks around 21 and 27 MeV in both the
 SIII and SGII results. The strength around $E_x$ = 21 MeV exhausts 
 50(41)\% of the NEWSR value, while the peak around $E_x$ = 27 MeV exhausts
 30(37)\% of the NEWSR value for the SIII (SGII) interaction. 
The peak energies in the two results are similar, while more SD strength is
shifted to the peak around $E_{x}$ = 21 MeV in the case of the SIII interaction.
The main configurations of the higher peak at $E_{x}$ = 27 MeV are the 
same as those of the 0$^{-}$ and 1$^{-}$ peaks, namely,
  $(\pi1h_{9/2}\nu1g_{9/2}^{-1})$ and  $(\pi1g_{7/2}\nu1f_{7/2}^{-1})$. On the other hand, the main configurations of 
the peak around $E_{x}$ = 21 MeV are 
 $(\pi1h_{11/2}\nu1g_{9/2}^{-1})$, 
$(\pi2d_{5/2}\nu2p_{1/2}^{-1})$ and $(\pi2d_{5/2}\nu2p_{3/2}^{-1})$.
The 2$^{-}$ strength distributions of the SkI3 and SLy4 interactions are 
 somewhat different
 than those of the SIII and SGII interactions. There is no isolated
low energy peak in the results for the SkI3 and SLy4 interactions. 
 Three large peaks are seen at $E_{x}$ = 23.5, 26.5 and 29.5 MeV 
together with several small peaks, while the two peaks
at 20 and 26 MeV exhaust most of the strengths in the case of SLy4. 

The SD strengths calculated by the 
  SD operator $\hat{S}^{\lambda}_{+ } =
 \sum_{i} t_{+}^{i}$ 
$r_{i}[\sigma \times Y_{1}(\hat{r}_{i})]^{\lambda}$
are shown in Fig. \ref{fig:zr90_sdp} (a), (b), (c) and (d) 
for the SIII, SGII, SkI3 and SLy4 interactions, 
respectively.  The strength distributions are divided into two energy regions: a broad bump below 10 MeV and a peak around $E_{x}$ = 13 MeV.
The strengths  below 10 MeV are due to the $\lambda^{\pi}$=1$^-$ and 2$^-$ states, 
while the high 
energy peak is induced mainly by the 1$^-$ states. The large  0$^-$ strength
is also found just above the high energy 1$^-$ peak. 
 The summed NEWSR values of all multipoles below 10 MeV are almost 
equal to the strength of the high energy peak around $E_{x}$ = 13 MeV in 
the case of the SIII and SGII interactions. The high energy peak of the 
 SIII interaction in Fig. 
 \ref{fig:zr90_sdp}(a) is 
about 2 MeV higher than those of
 SGII and 
SLy4, as listed in Table \ref{tab:Epeak}. The main
configuration for  the high energy peaks with  $\lambda^{\pi}$=0$^-$  1$^-$ and 2$^-$
is the $(\nu1g_{7/2}\pi1f_{7/2}^{-1})$ excitation. For the 
low energy 1$^-$ 
 strength, the $(\nu2d_{3/2}\pi2p_{3/2}^{-1})$ and 
 $(\nu1g_{7/2}\pi1f_{5/2}^{-1})$ configurations play the dominant roles.
 The $(\nu2d_{5/2}\pi2p_{1/2}^{-1})$,
 $(\nu2d_{5/2}\pi2p_{3/2}^{-1})$ and $(\nu3s_{1/2}\pi2p_{3/2}^{-1})$
 configurations have a large contribution in the low energy peak of 
$\lambda^{\pi}=2^-$.
 The $(\nu2d_{3/2}\pi1f_{7/2}^{-1})$ configuration contributes substantially 
 to the high energy 2$^-$ peak together with the 
 $(\nu1g_{7/2}\pi1f_{7/2}^{-1})$ configuration. The large spread in the distributions
of the SD 
strengths in Figs. \ref{fig:zr90_sdm} and \ref{fig:zr90_sdp} are 
due to the fact that the $p-h$ excitations are very different in 
unperturbed energy. Thus, 
the collision-less Landau damping effect plays an important role
  in the large
observed width of SD resonance, while the coupling to the continuum 
 plays a minor role. The coupling to 2-particle-2-hole (2p-2h) states 
 were shown to increase substantially the width of  the main peak of
 the $t_-$ SD excitations of $^{90}$Zr  in ref. \cite{Dro87}.

The energies of the main peaks E$_{peak}$ are 
tabulated in Table \ref{tab:Epeak} along with 
the average excitation energies, which are calculated by the
ratio of the energy-weighted sum rule (EWSR) $m_1$ to the non-energy 
weighted sum rule (NEWSR) $m_0$, \={E}$=m_1/m_0$. 
 The \={E} is always lower than 
 the E$_{peak}$ because of the low energy peak in the excitation 
spectra.
 For the $t_-$ response, the SkI3 interaction gives the highest excitation 
energy for the peak, while the SLy4 is the lowest.  Notice
that the energy of 
 SkI3 is the highest due  to the small effective mass $m^*/m$, while the negative 
Landau parameter $G_{0}'$ is responsible for the fact that SLy4 yields 
the lowest energy value in Table \ref{tab:Epeak}.
 The general trend of the average   
 excitation energy \={E} is the same for the $t_+$ response.  
The SIII, however, gives a somewhat 
 higher energy for the E$_{peak}$ than SkI3 does.

\begin{figure}[htp]
\includegraphics[width=4.5in,clip]{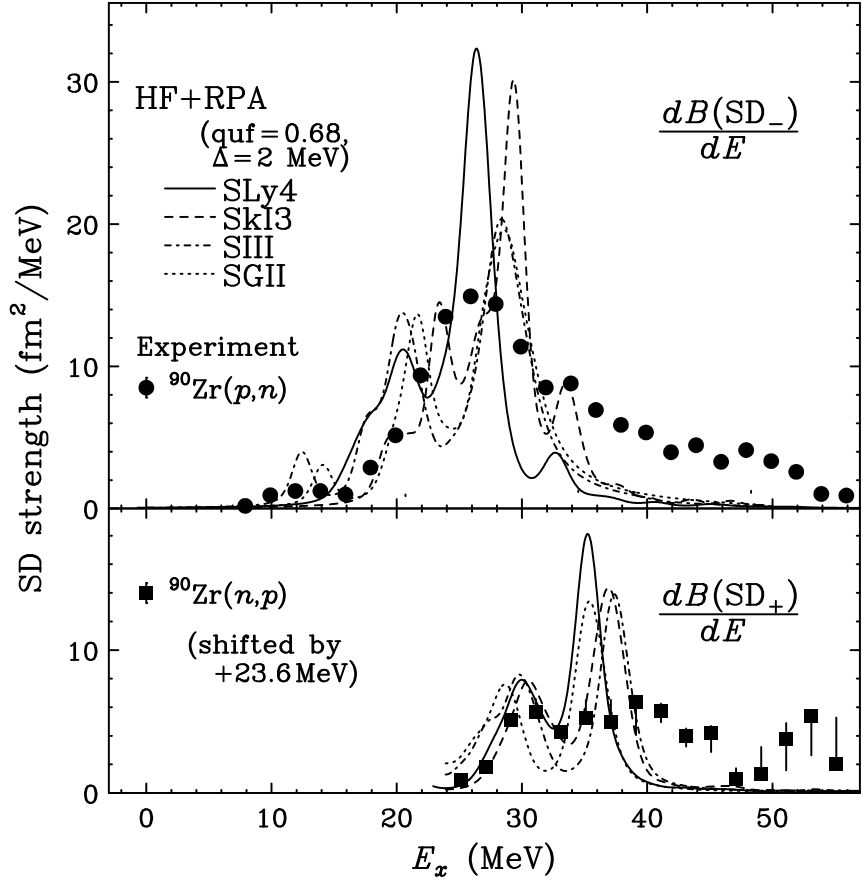}
\caption{\label{fig:zr90_sd_exp}
Charge exchange SD strength $\frac{dB(SD_-)}{dE}$ (upper panel) 
and $\frac{dB(SD_+)}{dE}$ (lower panel) of $^{90}$Zr.  The circles and squares 
are the experimental data taken from ref. \cite{Yako06}.
The spectra $\frac{dB(SD_+)}{dE}$ are shifted by the Coulomb
energy difference between the two daughter nuclei $^{90}$Nb and $^{90}$Y 
(+23.6 MeV) to adjust 
the isospin difference between the two nuclei.
The calculated results are plotted with the quenching factor
quf = 0.68. The SD strength is averaged by the weighting
 function (\ref{eq:weight}) with the width $\Delta $ = 2 MeV.}
 \end{figure}

The calculated results of SD strength are shown in Fig.~\ref{fig:zr90_sd_exp}
together with the experimentally obtained distributions of the
SD strengths \cite{Yako06}. The spectra for the $t_+$ channel are
shifted by +23.6 MeV, accounting for the Coulomb energy difference
between the daughter nuclei $^{90}$Nb and $^{90}$Yb. We introduce
 the quenching factor quf = 0.68 for both the $t_-$ and $t_+$ 
channels. For the $t_-$ channel,
 the experimental strength distribution peaked 
at $E_x\sim$ 26 MeV is well described by the SLy4 interaction. 
The results of SGII and SIII also give reasonable agreement with the
experimental peak energy.
None of the calculated results show any substantial strength above
 $E_x\sim$ 36 MeV, while a significant portion of the sum rule value 
is found above $E_x\sim$ 36 MeV in the experimental data. 
This difference may be due to the lack of coupling to many-particle
many-hole states in the present RPA calculations.
In ref. \cite{Dro87}, the $t_-$ SD strengths in $^{90}$Zr have been 
studied using the RPA model including the couplings to 
 2p-2h states. It was found that the 
  mixing between 1p-1h and 2p-2h 
states gives a large asymmetric spread in the strength of the 
SD resonances, and about 30$\%$ of the total strength is shifted to
excitation energies above 35 MeV, referred to the parent nucleus 
$^{90}$Zr. This result is consistent with the 
 quenching factor adopted in Fig. \ref{fig:zr90_sd_exp}. It should be mentioned that
 the peak energy of the $t_-$ SD strength is not changed
appreciably by the coupling to the 2p-2h states, while the peak height is decreased 
 substantially.

For the $t_+$ channel, 
the two peak structures can be seen 
in both the calculated and experimental results. 
  SkI3 and SLy4 describe 
the SD strength well at the low energy spectra. The calculated strength 
up to $E_x = $40 MeV exhausts 100\% of the sum rule value, 
while the experimental data show 
appreciable strength above $E_x = $40 MeV. 
This difference may be due to the couplings to many-particle many-hole 
states similar to the $t_-$ channel. 


\begin{figure}[htp]
\includegraphics[width=4.5in,clip]{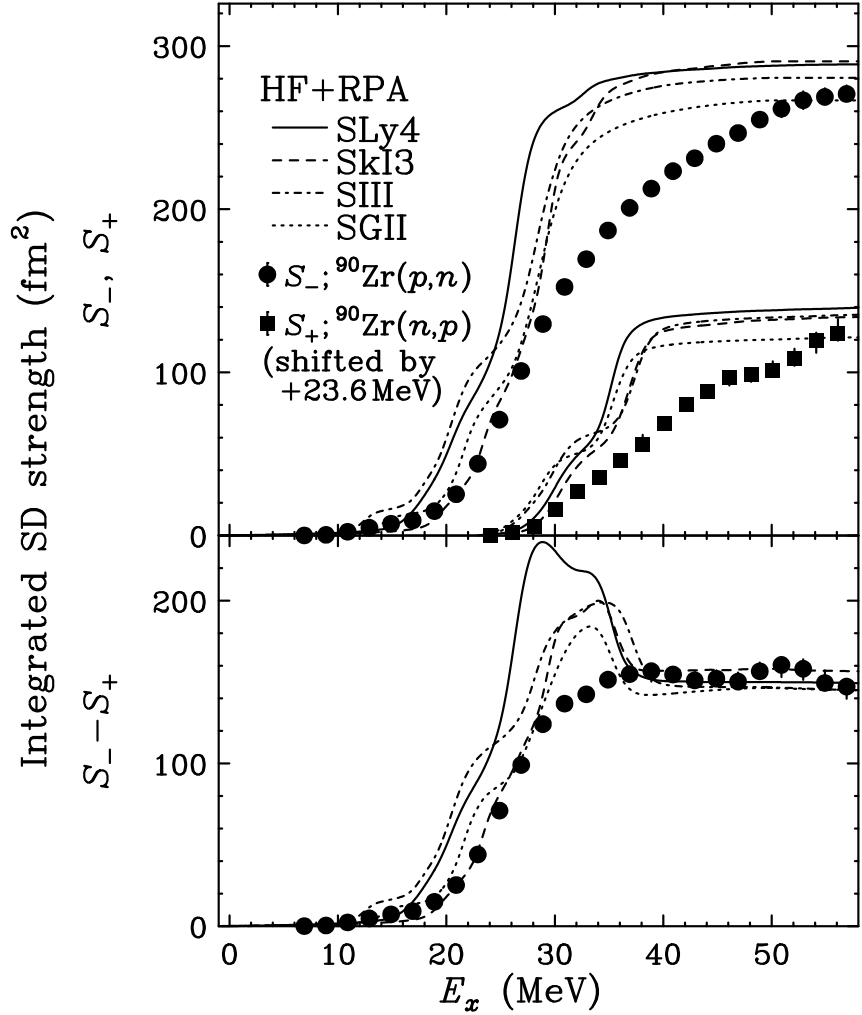}
\caption{\label{fig:zrsum}
Integrated charge exchange SD strength (\ref{eq:sum-ex}) 
excited by  the operators $
\hat{S}_{-} = \sum_{i,m,\mu} t_{-}^{i}\sigma_{m}^{i} r_{i}Y_{1}^{\mu}
(\hat{r}_{i})$ and  $
\hat{S}_{+} = \sum_{i,m,\mu} t_{+}^{i}\sigma_{m}^{i} r_{i}Y_{1}^{\mu}
(\hat{r}_{i})$ on $^{90}$Zr.  The
calculated results are obtained by the HF+RPA model using 
the Skyrme interactions  SIII,  SGII, SLy4 and  SkI3.
The upper panel shows the $S_-$ and $S_+$ strength, while 
the lower panel shows the  $S_--S_+$ strength.
 All strengths for the 
three multipoles $\lambda^{\pi}$=0$^{-}$, 1$^{-}$ and 2$^{-}$ are
summed up in the results.
The experimental data are taken from ref. \cite{Yako06}. 
 No quenching 
factor is introduced in the calculation of the integrated strength.
}
 \end{figure}

Let us now discuss the integrated SD strength.
The integrated SD strength 
\begin{equation}
 m_0(E_x)=\sum_{\lambda^{\pi}=0^-,1^-,2^-}\int_0^{E_x}\frac{dB(\lambda)}{dE'}
dE'
\label{eq:sum-ex}
\end{equation}
 is plotted as a function of 
the excitation energy $E_x$ in Fig. \ref{fig:zrsum} 
for the operators $\hat{S}^{\lambda}_{- } =
 \sum_{i} t_{-}^{i}$ 
$r_{i}[\sigma \times Y_{1}(\hat{r}_{i})]^{\lambda}$ 
and 
 $\hat{S}^{\lambda}_{+} =
 \sum_{i} t_{+}^{i}$ 
$r_{i}[\sigma \times Y_{1}(\hat{r}_{i})]^{\lambda}$.
The experimental data 
are taken from ref. \cite{Yako06}.
The value $S_-$ is obtained by integrating up to $E_x$ = 50 MeV from the 
ground state of the daughter nucleus $^{90}$Nb ($E_x$ = 57 MeV from 
the ground state of the parent nucleus $^{90}$Zr),
while the corresponding value $S_+$ is evaluated up to $E_x$ = 26 MeV 
from the ground state of $^{90}$Y ($E_x$ = 27.5 MeV from 
the ground state of the $^{90}$Zr). 
 This difference between the two maximum energies of the integrals 
 stems from the 
isospin difference 
  between the ground states of the daughter nuclei, i.e.,
 T=4 in $^{90}$Nb and T=6 in $^{90}$Y.
That is, the 23.6 MeV difference originates from the difference in 
excitation energy 
between the T=6 
  Gamow-Teller 
 states in the (p,n) and (n,p)
 channels\cite{Yako06}.
For both the $S_-$ and $S_+$ strength,
the calculated results overshoot the experimental data 
in the energy range $E_x$ = 20-40 MeV.  
  These results suggest the quenching of 
30-40\% of the calculated strength around the peak region, as was already 
mentioned.  
However, the integrated cross-sections up to $E_x$ = 56 MeV in Fig.
 \ref{fig:zrsum}  
approach the calculated values for both the 
$t_{-}$ and $t_{+}$ channels.

\begin{table}[htp]
\caption{\label{tab:sum}
  Sum rule values of charge exchange SD excitations in A=90 nuclei
obtained by the  HF+RPA calculations; S$_-$ for  $^{90}$Nb and
 S$_+$ for  $^{90}$Y.  The SD strength is integrated up to $E_x$ = 50 MeV 
for  S$_-$  and $E_x$ = 26 MeV for S$_+$, respectively. 
The experimental data are taken from ref. \cite{Yako06}.
The SD sum rules are given in units of fm$^2$.   
See the text for details.}
\footnotesize
\begin{tabular}{l||c|c|c||c|c|c||c|c|c||c|c|c} 
\hline
  &\multicolumn{3}{c||}{SIII} &\multicolumn{3}{c||}{SGII}
  &\multicolumn{3}{c||}{SkI3}&\multicolumn{3}{c}{SLy4} \\ \hline
$\lambda^{\pi} $ & $S_-$ & $S_+$ & $\Delta S$ & $S_-$ & $S_+$ & $\Delta S$
& $S_-$ & $S_+$ & $\Delta S$ & $S_-$ & $S_+$ & $\Delta S$ \\ \hline
 0$^-$  & 34.8  &18.5 &16.4 & 33.2 & 17.4 & 15.8  & 36.6 & 19.1& 17.5&37.8 &21.4 & 16.4\\\hline
 1$^-$  & 120.8 & 71.7 & 49.1 & 122.0 & 74.3 & 47.7 & 120.8  & 68.2 & 52.7 &115.8 & 66.4& 49.4\\\hline
 2$^-$  & 130.1 & 48.5 & 81.6 & 125.5 & 45.9  & 79.5 & 139.0 & 51.1 & 87.9&138.7 & 56.4& 82.3\\\hline
sum   & 285.7   & 138.6 &147.1  & 280.7  &137.6  &143.1   & 296.3 &  138.3   & 158.1 & 292.3 & 144.2 & 148.2\\ \hline \hline 
exp &  \multicolumn{3}{c||}{ $S_-=271\pm14$} &\multicolumn{3}{c||}{
$S_+=124\pm11$ }
  &\multicolumn{3}{c||}{$\Delta S=147\pm13$}&\multicolumn{3}{c}{} \\ \hline
\end{tabular}
\end{table}

The  calculated SD sum rule values in A=90 nuclei obtained by using the HF+RPA results 
 are  tabulated in Table \ref{tab:sum} for the transitions with $\lambda ^{\pi}$=
 0$^-$, 1$^-$ and 2$^-$. 
Clearly, the $\Delta S$ values show
signs of multipole proportionality (2$\lambda$+1), 
 even though $S_-$ and $S_+$ themselves 
 do not show any clear multipole dependence. 
The present 
RPA results for $^{90}$Zr listed in Table \ref{tab:sum} 
 satisfy the sum rule value (2) 
  in Table \ref{tab:hf-zr}
  with high accuracy, to an error of only (0.1$\sim$0.2)\%. 
 This agreement guarantees the numerical 
  accuracy of the present RPA calculations. 
 This is also the case in $^{208}$Pb, as will be shown in Section IIB. 
The $\Delta S=S_--S_+$ value is shown as a function of $E_x$ 
 in the lower panel of Fig. 
 \ref{fig:zrsum}.  We note that the 
 $\Delta S$ value saturates both in the calculated and the experimental 
 values above $E_x=$ 40 MeV, while the empirical values $S_-$ and $S_+$
 themselves increase gradually above $E_x=$ 40 MeV.  
This is the crucial feature for extracting the model-independent sum rule 
$\Delta S=S_--S_+$ from the experimental data.
The empirical values $S_-$,  $S_+$ and  $\Delta S$ obtained from these 
analyses are shown in Table \ref{tab:sum}. 
   The indicated uncertainties of $S_-$, $S_+$ and $\Delta S$ contain not only
   the statistical error of the data, but also errors due to the various
   input of the DWIA calculations used in the MD analysis, such as, 
   the optical model parameters and the single-particle potentials
   \cite{Yako}. There is an additional uncertainty in the 
   estimation of the SD unit cross-section, namely, the overall 
   normalization factor  \cite{Yako06}, which should be studied further 
   experimentally.
 
 From  $\Delta S$, the neutron radius of
$^{90}$Zr is extracted to be $\sqrt{<r^2>_n}$ = (4.26$\pm$0.04)fm 
from the model-independent SD sum rule (\ref{eq:sum_sd}), where the empirical
proton radius $\sqrt{<r^2>_p}$ = 4.19 fm is used. The proton radius 
is obtained from the charge radius in Table \ref{tab:hf-zr} by
 subtracting the
proton finite size correction. 
The experimental uncertainty in the neutron skin thickness 
obtained by proton scattering is rather large: 
   $\delta_{np}=r_n-r_p=(0.09\pm0.07$)fm.  This is 
 because of the model-dependent
analysis of the proton scattering, with effective nucleon-nucleon 
interactions in the nuclei \cite{pscatt1}.
On the other hand, 
 the sum rule analysis of the SD strength 
determines the neutron radius with 1\% accuracy, which is 
almost the same as  that expected for  the parity violation electron 
 scattering experiment.
 The obtained value  $r_n-r_p = (0.07\pm0.04$) fm
  can  be used to disentangle the neutron matter 
 EOS by using the strong linear correlation between the two quantities 
  \cite{Brown,Furn,Yoshi04}, as will be discussed in Section 3.

\subsection{Charge exchange SD excitations of  $^{208}$Pb}

\begin{table}[htp]
\caption{\label{tab:hf-pb}
 Proton, neutron and charge radii of $^{208}$Pb.
 The charge radius is obtained by folding the proton finite size.
 The sum rule values $\Delta S=S_--S_+$ of the SD excitations are calculated 
 by Eq. (\ref{eq:sum_sd}) with the HF neutron and proton mean square radii.  
Experimental data on the charge radius are taken from ref. \cite{Vries}.
Experimental data on $\delta_{np}=r_n-r_p$ are obtained by the 
proton scattering \cite{pscatt2,Hoffman,Starodubsky} and the 
giant dipole excitations of $^{208}$Pb\cite{GDR}. The radii are given in 
units of fm, while the SD sum rules are given in units of fm$^2$.
}

\begin{tabular}{l|c|c|c|c|l}  
 \hline
          & SIII  & SGII  & SkI3   & SLy4  & exp \\\hline
$r_p$     & 5.521 & 5.454 & 5.421  & 5.457 & ---    \\
$r_c$     & 5.578 & 5.512 & 5.479  & 5.515 & 5.503 $\pm$ 0.002    \\
$r_n$     & 5.646 & 5.589 & 5.649  & 5.617 & ---    \\ \hline
$\delta_{np}=r_n-r_p$ & 0.125 & 0.135 & 0.228  & 0.160 &  $0.083 < \delta_{np} < 0.111$\cite{pscatt2}, $0.19 \pm0.09$\cite{GDR}     \\
\hline
 $\Delta S$ & 1086. & 1072. & 1154. & 1098. & \\ \hline
\end{tabular}
\end{table}


\begin{figure}[htp]
\includegraphics[width=3.2in,clip]{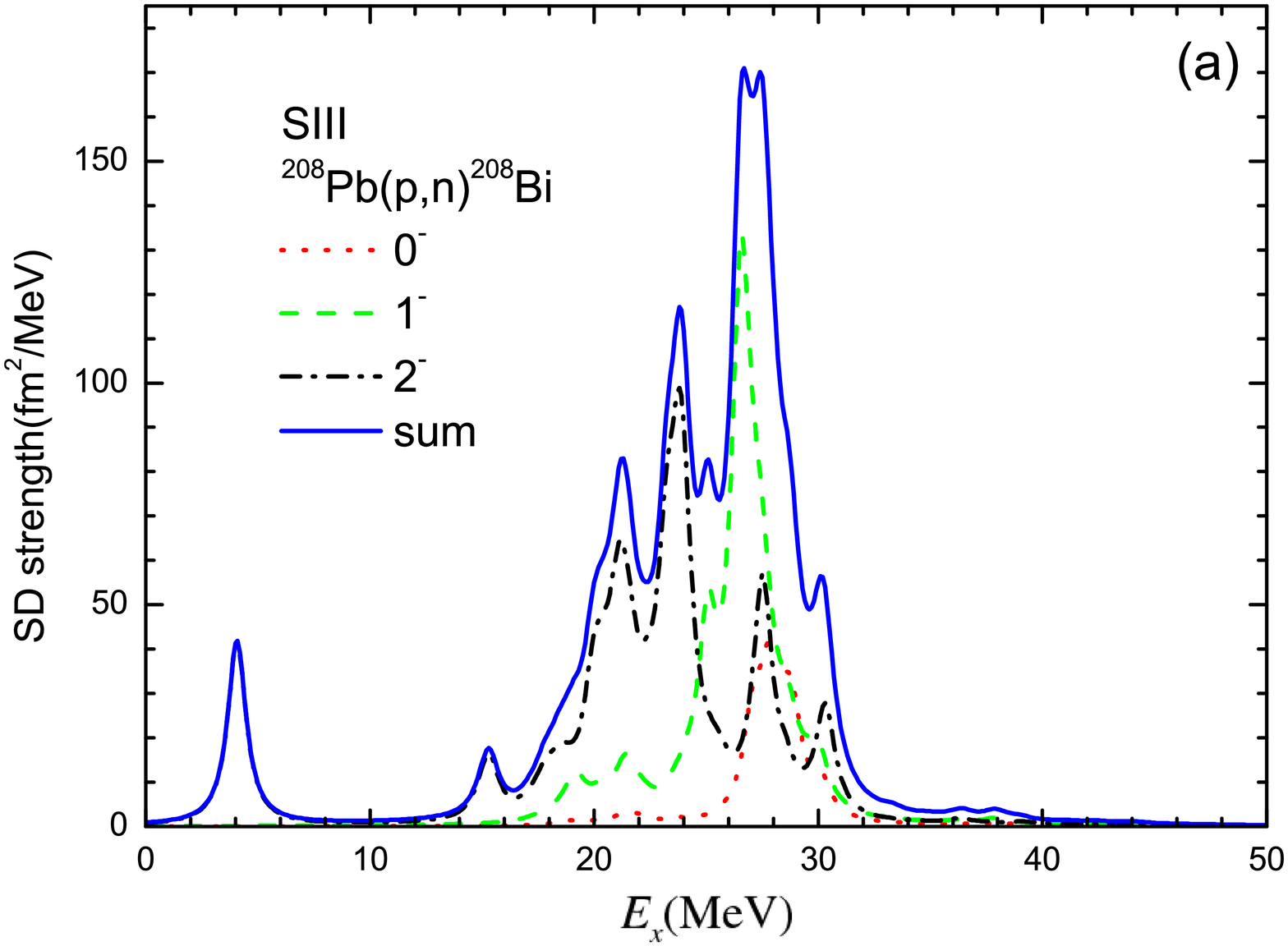}
\vspace{1cm}
\includegraphics[width=3.2in,clip]{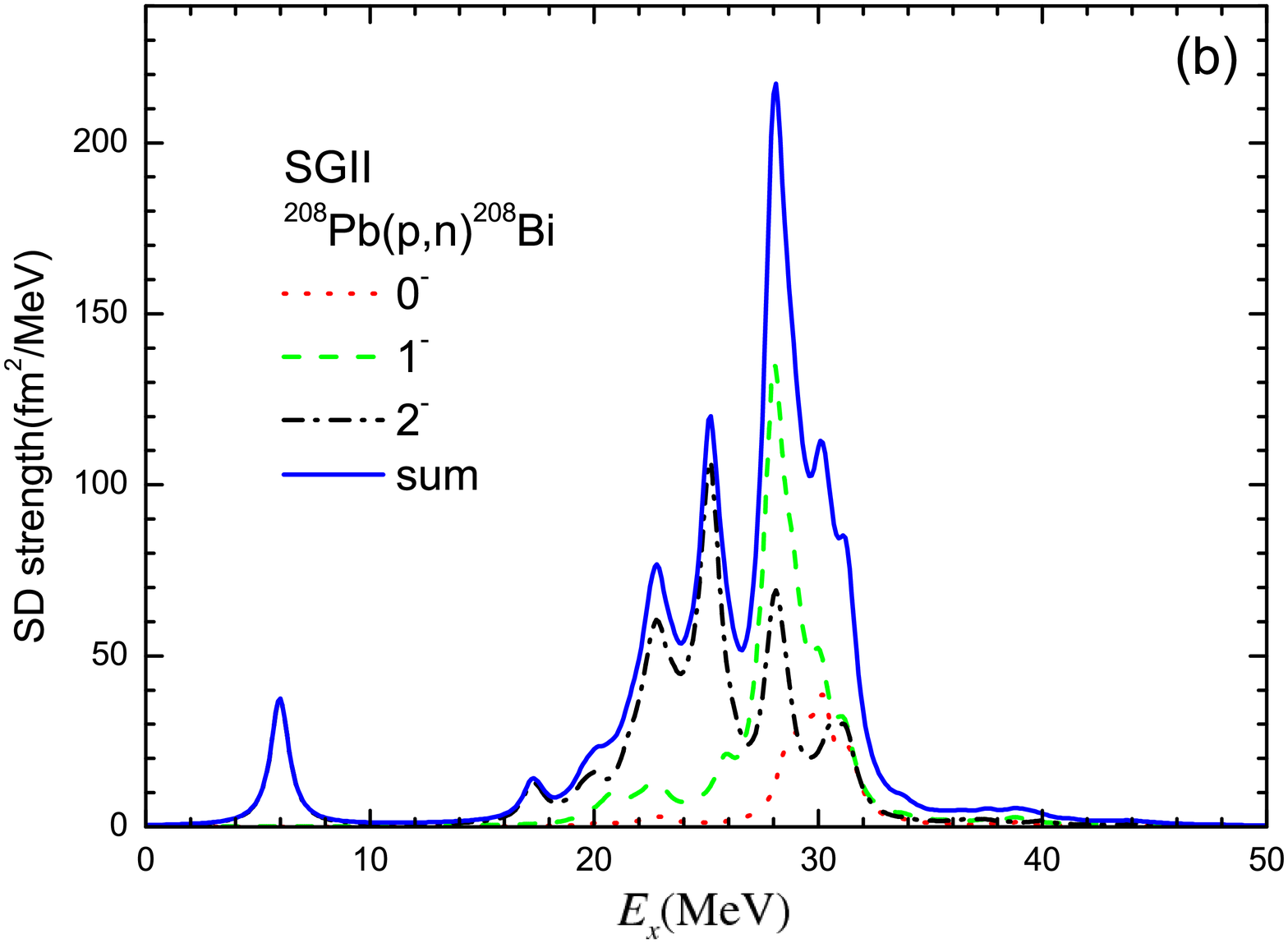}
\includegraphics[width=3.2in,clip]{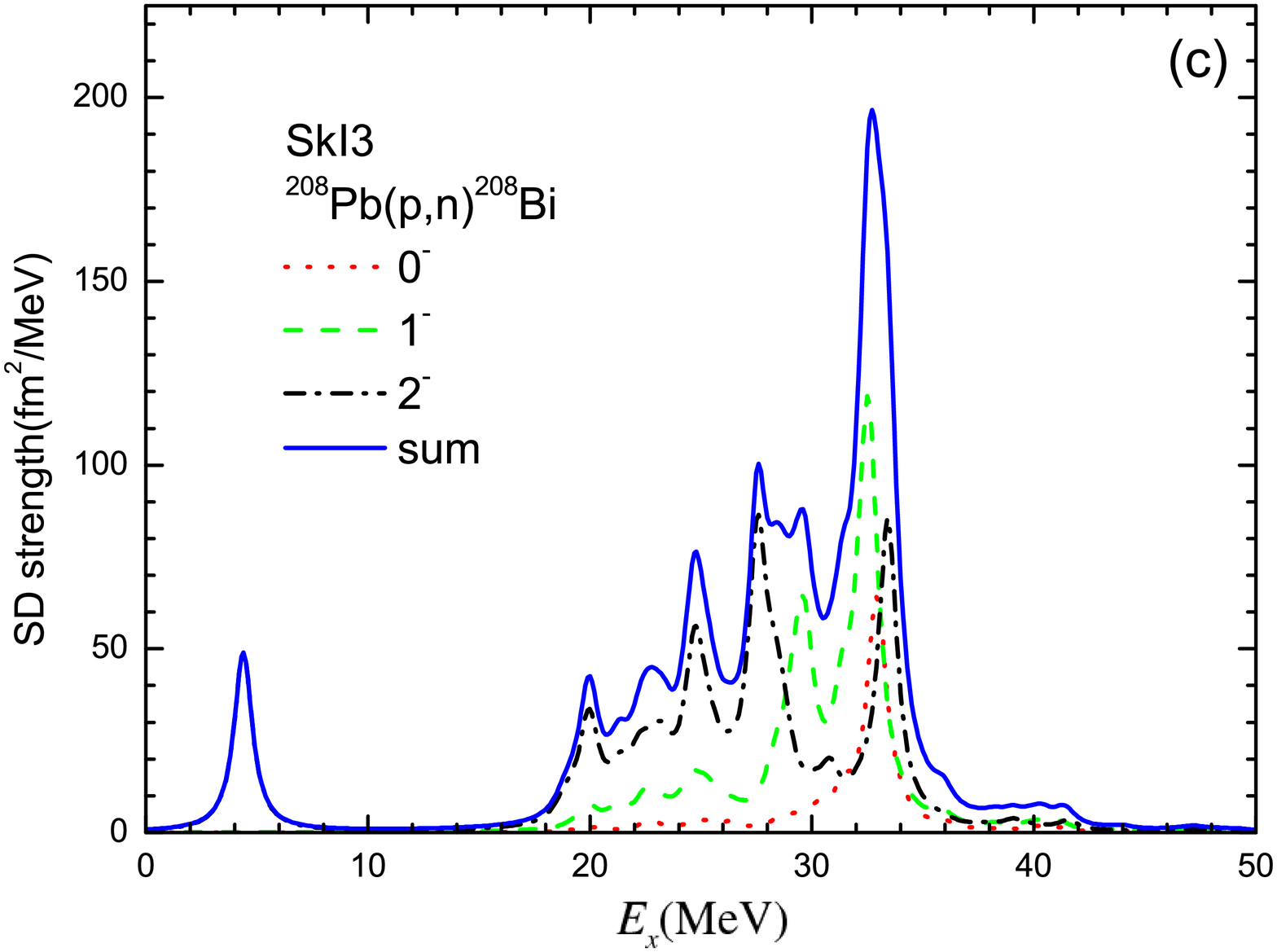}
\includegraphics[width=3.2in,clip]{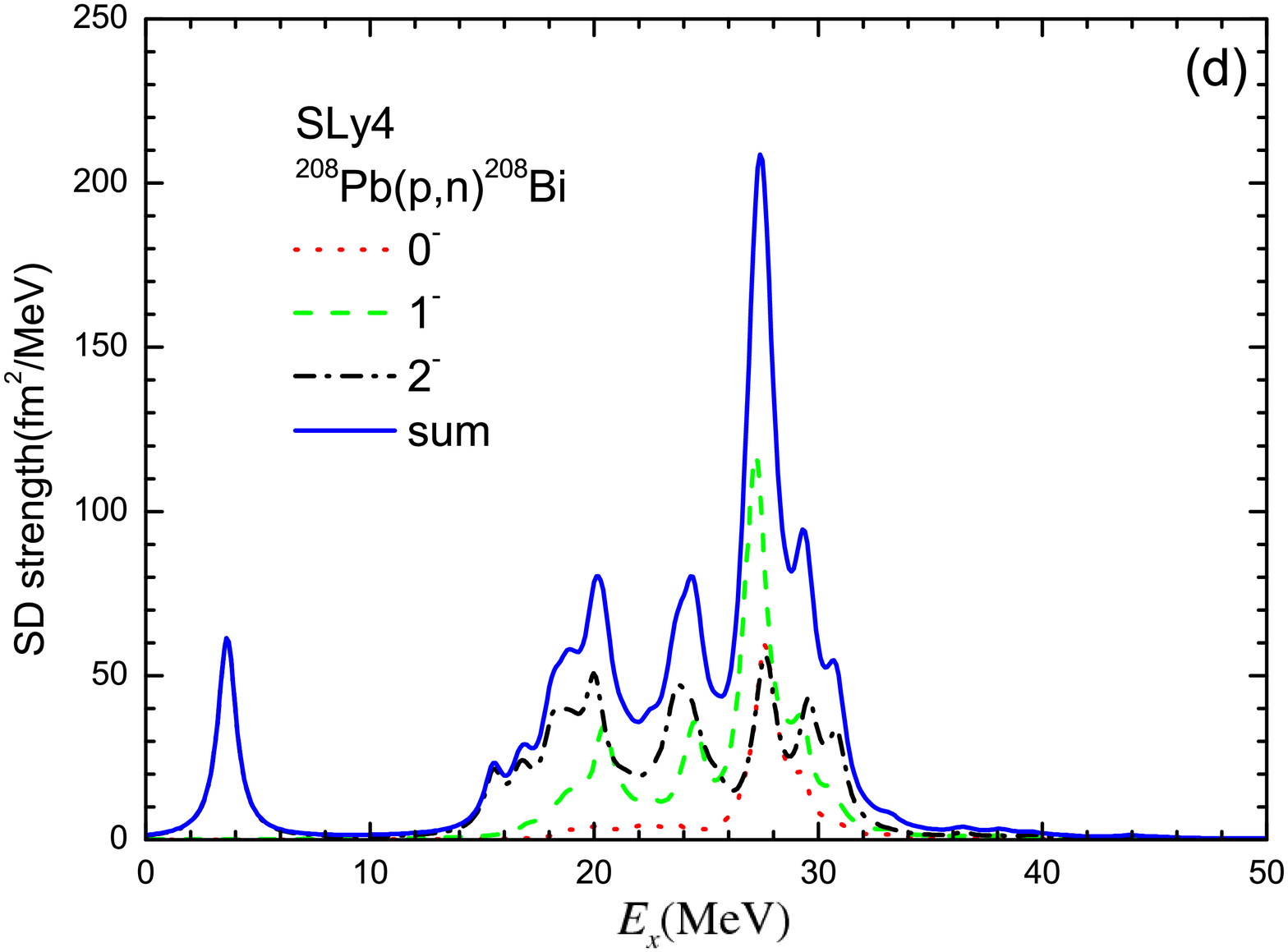}
\caption{\label{fig:pb208_sdm}(Color online)
Charge exchange SD strengths for  the operators  $\hat{S}^{\lambda}_{-} =
 \sum_{i} t_{-}^{i}$ 
$r_{i}[\sigma \times Y_{1}(\hat{r}_{i})]^{\lambda}$
calculated by the  HF+RPA model with 
the Skyrme interactions (a) SIII, (b) SGII, (c) SkI3, and (d) SLy4.  
The excitation energy is referred to the ground state of the
parent nucleus $^{208}$Pb. 
The SD strength is averaged by the weighting function in Eq. (\ref{eq:weight})
  with the width $\Delta $ = 1 MeV.}
 \end{figure}

\begin{figure}[htp]
\includegraphics[width=3.2in,clip]{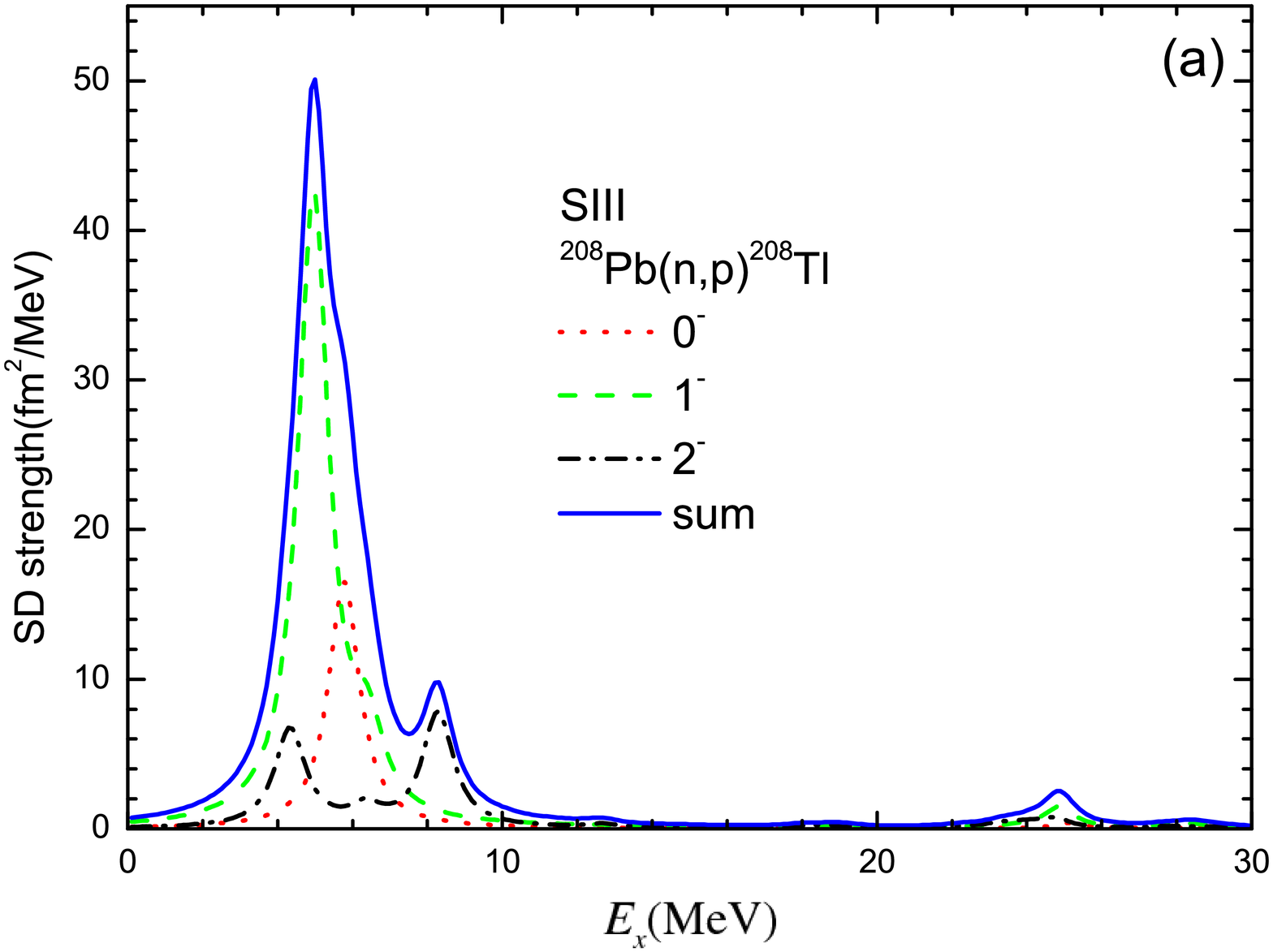}
\vspace{1cm}
\includegraphics[width=3.2in,clip]{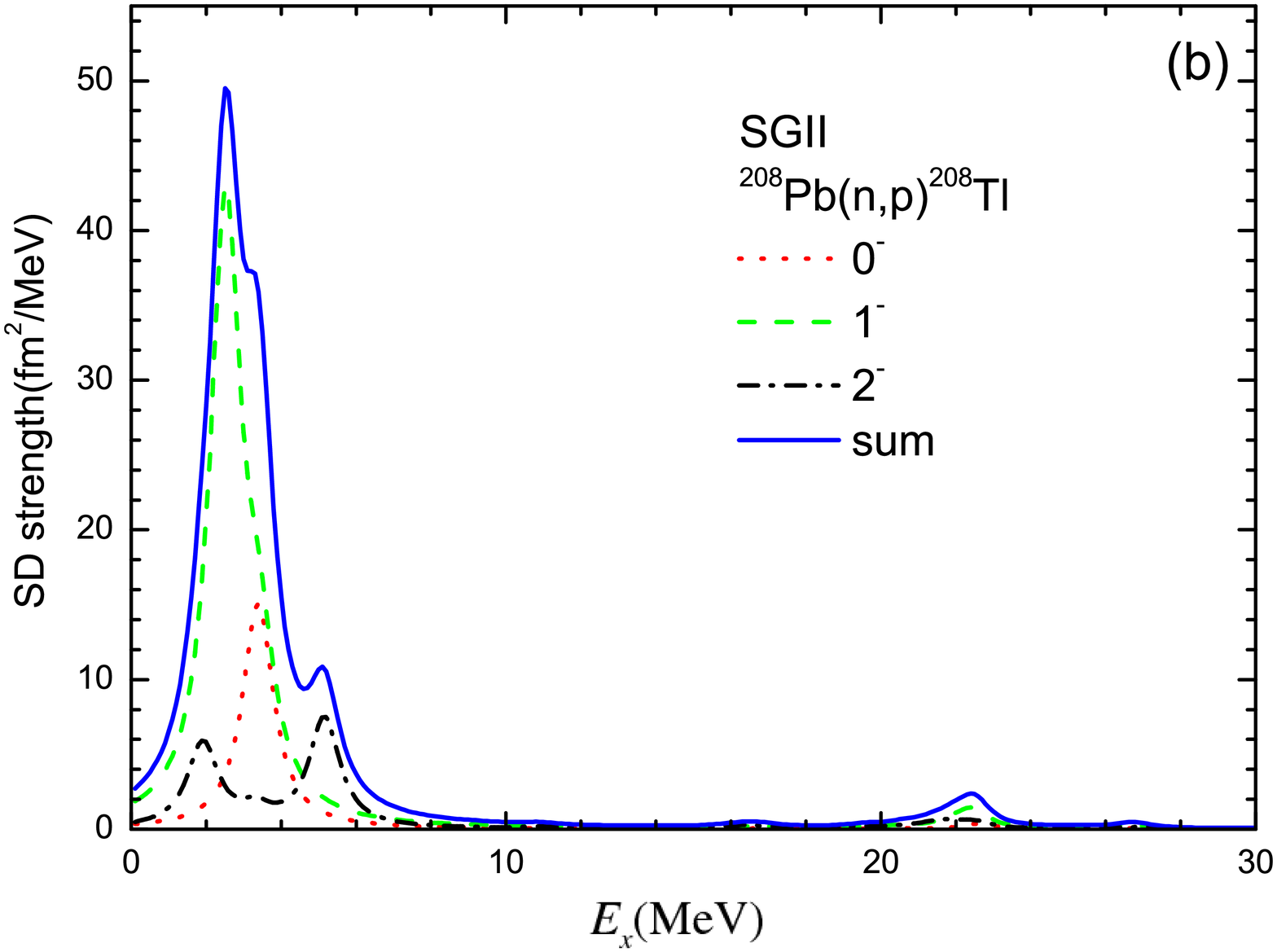}
\includegraphics[width=3.2in,clip]{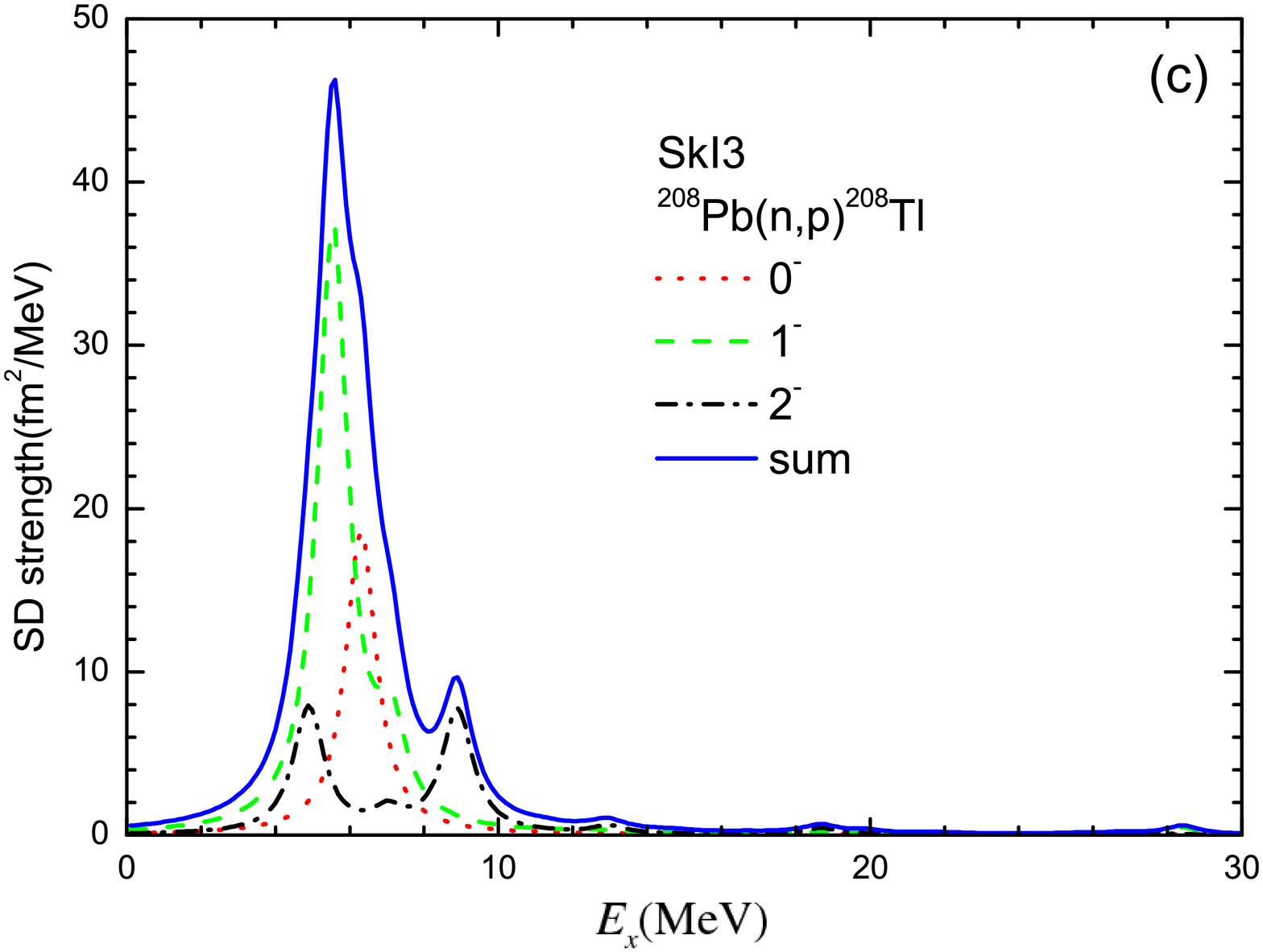}
\includegraphics[width=3.2in,clip]{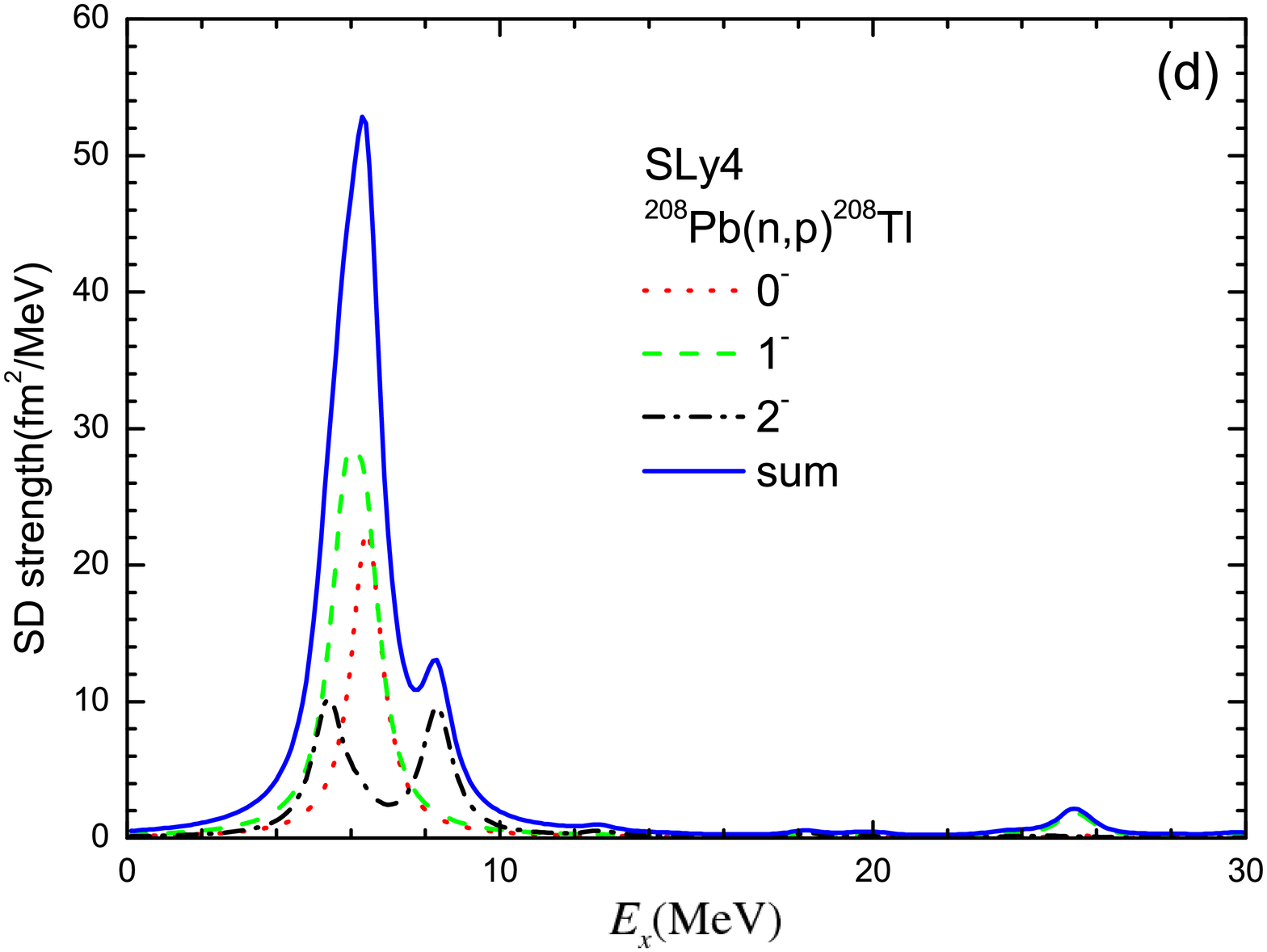}
\caption{\label{fig:pb208_sdp}(Color online)
Charge exchange SD strengths for  
 the operators  $\hat{S}^{\lambda}_{+} =
 \sum_{i} t_{+}^{i}$ 
$r_{i}[\sigma \times Y_{1}(\hat{r}_{i})]^{\lambda}$ 
calculated by the  HF+RPA model with 
the Skyrme interactions (a) SIII, (b) SGII, (c) SkI3 and (d) SLy4.
The excitation energy is referred to the ground state of the
parent nucleus $^{208}$Pb. 
The SD strength is averaged by the weighting function in 
  Eq. (\ref{eq:weight}) with the width
$\Delta$ = 1 MeV.}
 \end{figure}

\begin{figure}[htp]
\includegraphics[width=4.in,clip]{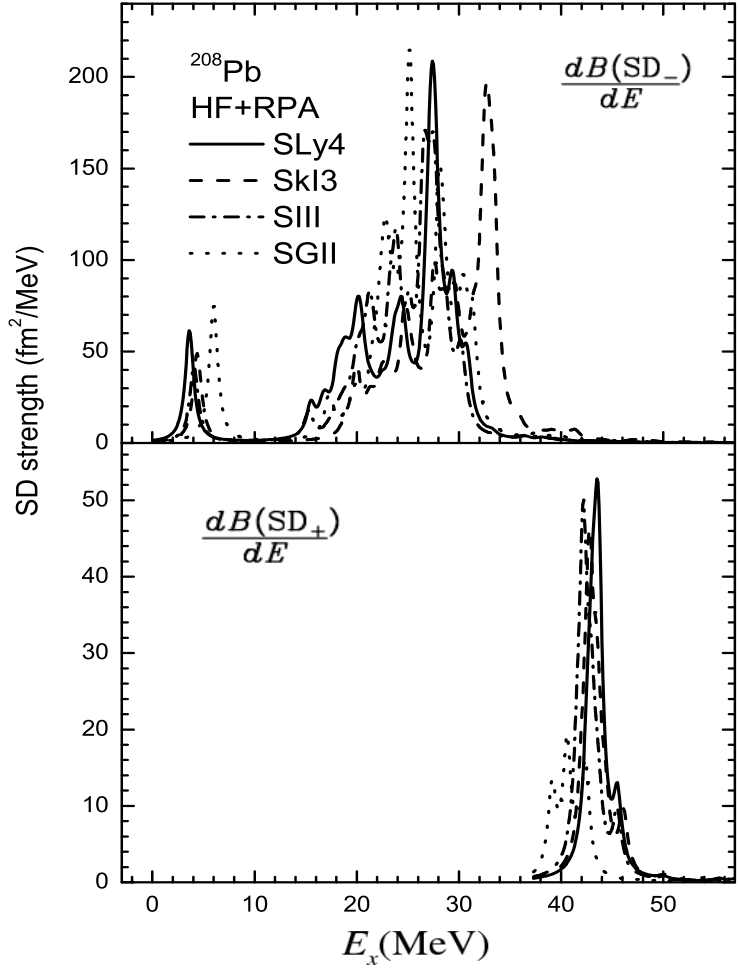}
\caption{\label{fig:pb_sdmp}
Charge exchange SD strength $\frac{dB(SD_-)}{dE}$ (upper panel) 
and $\frac{dB(SD_+)}{dE}$ (lower panel) of $^{208}$Pb.  
 The spectra $\frac{dB(SD_+)}{dE}$ are shifted by +37.2 MeV due to
the Coulomb energy difference between the two daughter nuclei $^{208}$Bi and 
$^{208}$Tl. The arrow in the upper panel 
shows a peak energy at $E_x$ = 24.8 MeV 
 observed by the charge exchange reaction 
 $^{208}$Pb($^3$He,$t$)$^{208}$Bi \cite{Aki}.  
}
 \end{figure}

\begin{figure}[htp]
\includegraphics[width=4.in,clip]{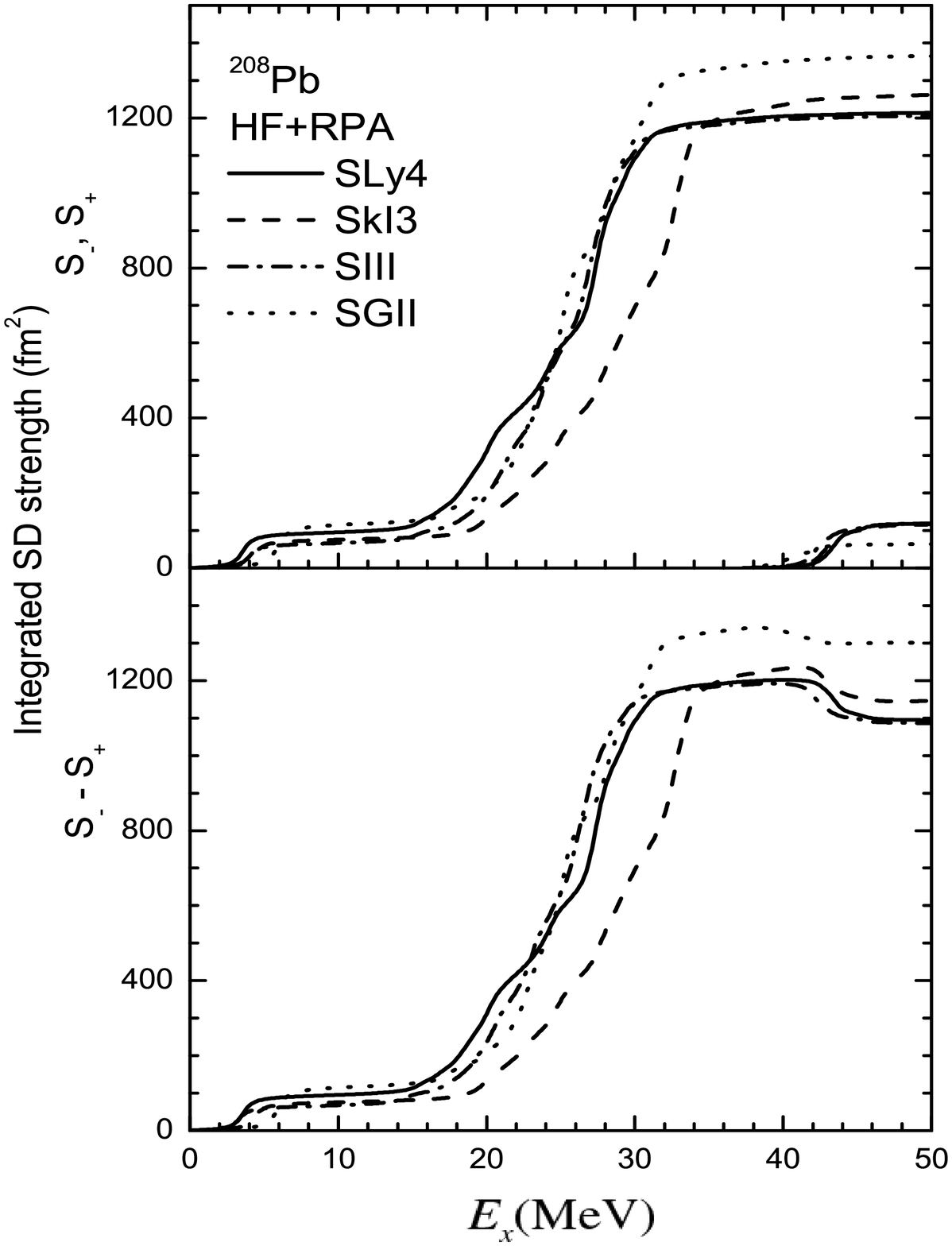}
\caption{\label{fig:pbsum}
Integrated charge exchange SD strength (\ref{eq:sum-ex}) of $^{208}$Pb
for the operators $
\hat{S}_{-} = \sum_{i,m,\mu} t_{-}^{i}\sigma_{m}^{i} r_{i}Y_{1}^{\mu}
(\hat{r}_{i})$ and  $
\hat{S}_{+} = \sum_{i,m,\mu} t_{+}^{i}\sigma_{m}^{i} r_{i}Y_{1}^{\mu}
(\hat{r}_{i})$
calculated by the HF+RPA model with 
the Skyrme interactions SIII, SGII, SkI3 and SLy4.
The upper panel shows the $S_-$ and $S_+$ strength, while 
the lower panel shows the  $\Delta S=S_--S_+$ strength.
 All strengths for the 
three multipoles $\lambda^{\pi}$=0$^{-}$, 1$^{-}$ and 2$^{-}$ are
summed up in the results.
}
 \end{figure}

The HF results of $^{208}$Pb are summarized  in Table
  \ref{tab:hf-pb}.  
The RPA results of SD excitations of $^{208}$Pb are given
 in Figs. \ref{fig:pb208_sdm} and \ref{fig:pb208_sdp} 
 for the four different Skyrme 
interactions, namely, SIII, SGII, SkI3, and SLy4.  
For the $t_-$ channel, the strength distributions are spread out in a broad
 energy region (15 MeV$<E_x<$35 MeV) except a tiny peak 
at $E_x\sim$ 5 MeV. On the other hand, the strength for the  $t_+$ channel is 
concentrated in a single narrow peak.
 The highest peak of the $t_-$ channel
occurs at $E_x\sim$ 27-28 MeV in the cases of the
SIII, SGII, and SLy4 interactions, while it is shifted to
higher energies ($E_x\sim$ 33 MeV) in the case of SkI3.
  The 0$^-$ and   1$^-$ excitations    
 have merged into one 
peak, 
  having  more than 40\% of the total strength at the high energy side, 
 while the 2$^-$ states split into a broad energy region.  
 The low-energy 2$^-$ state at around $E_x$ = 4 MeV  
   is mainly due to the ($\pi1h_{9/2}\nu1i_{13/2}^{-1}$)
excitation. 
  The 0$^-$ peak is predicted to occur at a slightly higher energy than the 1$^-$
 peak. However, 
  it might be difficult to observe this peak experimentally 
 because 
of its rather low strength. 
 There are appreciable differences in the peak energies between the Skyrme 
interactions 
 for the $t_+$ channel: 
  $E_x\sim$ 3 MeV for SGII,  $E_x\sim$ 5 MeV for SIII and  $E_x\sim$ 6 MeV 
for SLy4 and SkI3, as listed in Table \ref{tab:Epeak-pb}.
 The sum rule values  $S_-$ and  $S_+$  are listed in Table
  \ref{tab:sum-pb}.  Because of the strong Pauli 
  blocking of neutron excess
  in $^{208}$Pb, the $S_+$ value is much smaller than the  S$_-$ value,
  at most,  20\% of the corresponding $S_-$ value for each multipole.
 The  $S_+$ value is substantial in the case of A=90 
  as  shown in Table  \ref{tab:sum}, more 
than 55\% of $S_{-}$ in some cases. However, $\Delta S=S_--S_+$ obeys
the (2$\lambda $+1) proportionality, as expected from Eq. (\ref{eq:sum_sd_a}).  
The charge exchange $^{208}$Pb($^3$He,$t$)$^{208}$Bi reaction
was performed to study the SD strength in $^{208}$Bi.
The data were analyzed by a least-squares fitting method and the peak 
of the SD strength was  found  to be at $E_{x}$ = 24.8$\pm$0.8 MeV, as measured 
from the ground state of $^{208}$Pb \cite{Aki}. This empirical
peak energy is close to the average energy $\bar{E}$ of SD strength
obtained by SIII and SGII in 
Table \ref{tab:Epeak-pb}.  Further experimental effort is 
urgently needed to obtain more quantitative strength 
distributions, for example, for the multipole decomposition analysis 
of charge exchange reactions on a $^{208}$Pb target.
 
\begin{table}[htp]
\caption{\label{tab:Epeak-pb}
  Peak energies and the average energies 
 of charge exchange SD excitations in A=208 nuclei calculated 
by the HF+RPA model; S$_-$ for $^{208}$Bi and
 S$_+$ for  $^{208}$Tl. The average energy is calculated 
by the ratio of EWSR to NEWSR: \=E(MeV)=$m_1/m_0$. See the text 
for details }
\footnotesize\rm
\begin{tabular}{l|c|c|c|c} 
\hline
  &\multicolumn{2}{c|}{$t_-$} &\multicolumn{2}{c}{$t_+$} \\ \hline
   & $E_{peak}$(MeV) & \=E(MeV) &  $E_{peak}$(MeV) & \=E(MeV) \\ \hline
 SIII & 26.7  & 24.2   &  5.0    &  7.3  \\
 SGII & 28.1   & 24.6   &   2.5   &    6.0  \\
 SkI3 & 32.7 &  27.9   &  5.6  &    7.3   \\
  SLy4 & 27.4  &  23.6  &   6.3    &  8.0  \\ \hline
\end{tabular}
\end{table}

\begin{table}[htp]
\caption{\label{tab:sum-pb}
  Sum rule values of charge exchange SD excitations in A=208 nuclei
 calculated by the HF+RPA model; S$_-$ for $^{208}$Bi and
 S$_+$ for  $^{208}$Tl.  The SD strength is integrated up to $E_x$ = 57 MeV 
for  S$_-$   and $E_x$ = 20 MeV for S$_+$; the excitation energy is referred 
to the ground state of  $^{208}$Pb.  The SD sum rules are given in units of fm$^2$.  
See the text for details.}
\footnotesize\rm
\begin{tabular}{l|c|c|c|c|c|c|c|c|c|c|c|c} 
\hline
  &\multicolumn{3}{c|}{SIII} &\multicolumn{3}{c|}{SGII}
  &\multicolumn{3}{c|}{SkI3}&\multicolumn{3}{c}{SLy4} \\ \hline
$\lambda^{\pi} $ & $S_-$ & $S_+$ & $\Delta S$ & $S_-$ & $S_+$ & $\Delta S$
& $S_-$ & $S_+$ & $\Delta S$ & $S_-$ & $S_+$ & $\Delta S$ \\ \hline
 0$^-$  & 148.6  & 27.0 &121.6& 114.1 & 24.3 & 119.8  & 158.0 & 29.7 & 128.3&158.5 & 36.0  & 122.5\\\hline
 1$^-$  & 442.7 & 78.8 & 363.9& 440.4 & 82.3 & 358.1 & 454.5 & 69.2 & 385.3 &430.8& 63.6 & 367.2\\\hline
 2$^-$  & 632.2 & 28.3& 603.9 & 620.7 & 26.4  & 595.3 & 669.8 & 28.2 & 641.6&644.5 & 34.1 & 610.5\\\hline
sum   & 1224.  & 134.1 & 1089.  & 1205.  &132.0  & 1073.  & 1282. &  127.1   & 1155.  & 1234. & 133.7 & 1100. \\ \hline 
\end{tabular}
\end{table}

One can see only one sharp peak in the $t_+$ channel in Fig. 
 \ref{fig:pb208_sdp}.
There are only two allowed 1p-1h configurations 
($\nu2g_{9/2}\pi1h_{11/2}^{-1}$) and ($\nu1i_{11/2}\pi1h_{11/2}^{-1}$)
 for both 1$^-$ and 2$^-$ excitations because of the strong 
Pauli blocking effect of excess neutrons. Moreover, the 
  $\nu2g_{9/2}$  and  $\nu1i_{11/2}$ states are almost degenerate 
 in energy in the HF potential. They are the reasons why there is only 
one sharp peak in the $t_+$ channel of $^{208}$Pb. 
 It might be interesting to perform $^{208}$Pb(n,p)$^{208}$Tl or
  $^{208}$Pb($t,^3$He)$^{208}$Tl
reactions in order to observe this peak experimentally.
The $^{208}$Pb(n,p) $^{208}$Tl reaction has been reported for 
the $t_+$ channel, and a broad peak found at $E_{x}\sim$ 8MeV, 
as measured from the ground 
state of $^{208}$Pb with rather poor statistics \cite{Long}.

The integrated SD strengths for both the $t_-$ and $t_+$ channels 
are shown in Fig. \ref{fig:pbsum}.  The calculated NEWSR shows a
saturation at around $E_x\sim$ 30 MeV as can be seen in  Fig. \ref{fig:pbsum}.
As noted previously, the $t_+$ channel has only a small contribution 
to the model-independent sum rule $\Delta S$.  

The couplings to the 2p-2h states may increase the spread in the SD strength in A=208 nuclei as well as A=90 nuclei.
So far, the charge exchange Gamow-Teller(GT) states in $^{208}$Bi 
were studied by taking into account the couplings to 2p-2h states in 
the particle-vibration model \cite{Colo}. While a large spread was found in the 
GT states in the particle-vibration model calculations, 
 the peak energy did not change appreciably due to the couplings to 
  2p-2h states. 
    There have been no microscopic studies of SD states that take 
 into account the couplings to 2p-2h states in A=208 nuclei.

\section{SD sum rules and neutron matter EOS}
 
Sum rules are useful tools to study the collective nature of excitation modes
in many-body systems. In particular, for charge exchange excitations,
model-independent sum rules
 are derived and used to analyze experimental 
data on Gamow-Teller resonances and SD resonances \cite{Gaarde}. 
For SD states, the sum rules can be used to extract the 
neutron skin thickness, as was discussed in Section 2.
References \cite{Brown,Furn,Yoshi04} have reported a strong correlation
between the neutron skin thickness and the neutron matter EOS, as obtained
by using Skyrme and relativistic mean field theories.
In this section, we will  study  the relation between 
the SD sum rules   and  the neutron matter EOS.
The strong linear correlation between the neutron skin thickness
\begin{equation}
\delta_{np}=\sqrt{\langle r^2\rangle _n}-\sqrt{\langle r^2\rangle _p}
\end{equation}
and the pressure of neutron matter
\begin{equation}
P=\rho_n \frac{d(E(\rho_n)/\rho_n)}{d\rho_n}.
\end{equation}
is essential for this study.  
 Other linear correlations between the 
neutron skin thickness and various isovector nuclear matter 
properties have also been pointed out recently \cite{Yoshi06}. Given these 
correlations, accurate information on the neutron skin thickness
will be quite useful in determining empirically the pressure of neutron
   matter EOS
and 
 isovector nuclear properties,
such as the volume and surface symmetry energies.    

\begin{figure}[htp]
\includegraphics[width=4.5in,clip]{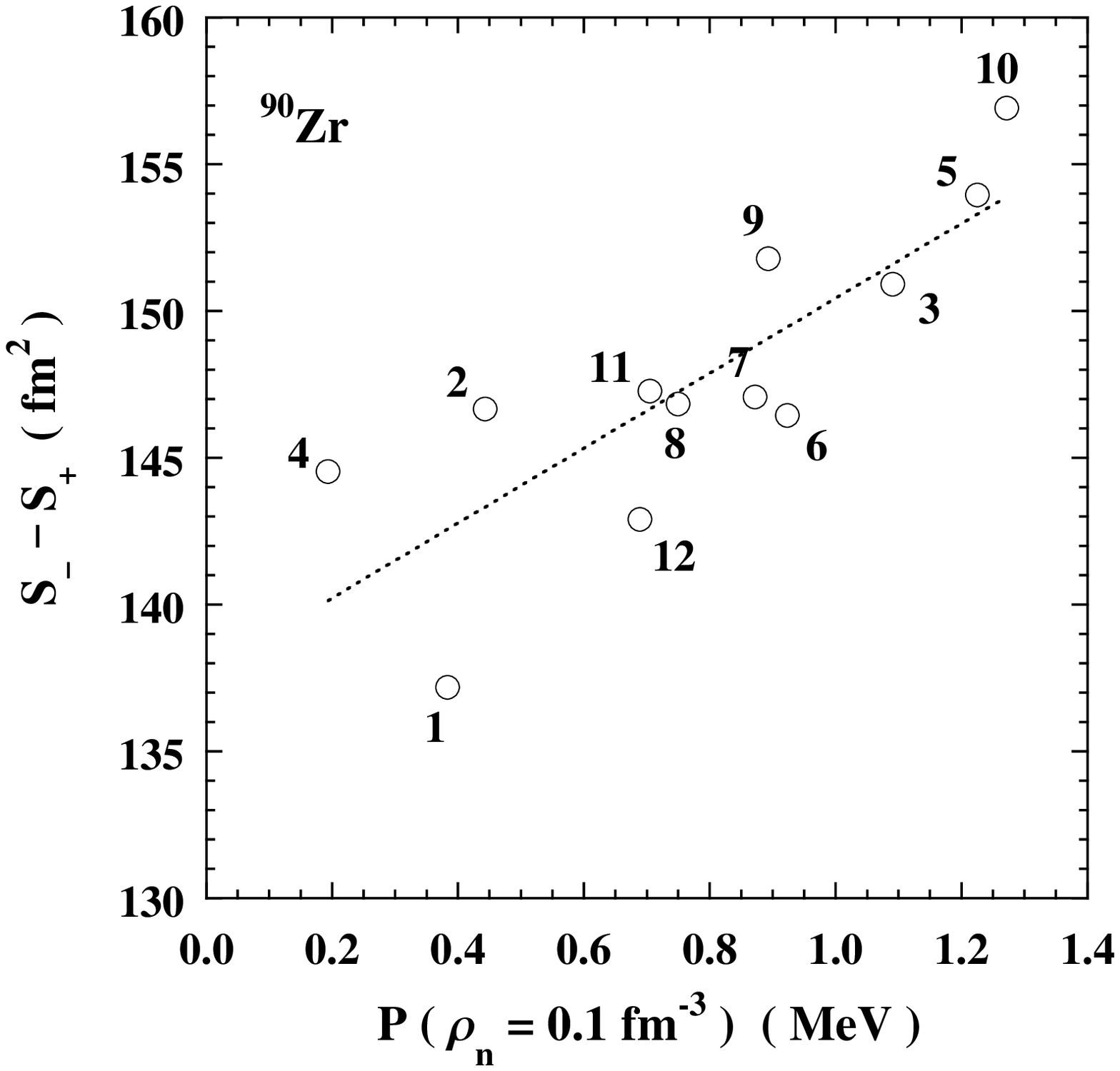}
\caption{\label{fig:zr90_np}
Correlations between the pressure of neutron matter and the SD sum rule 
values of $^{90}$Zr with 12 different Skyrme interactions. The numbers denote 
  different Skyrme parameter sets: 1 for SI, 2 for SIII, 3 for 
SIV, 4 for SVI, 5 for Skya, 6 for SkM, 7 for SkM$^{*}$, 8 for SLy4, 9 for MSkA, 
10 for SkI3, 11 for SkX and 12 for SGII.
The correlation coefficient
 is found to be r = 0.811.  }
 \end{figure}

\medskip
\medskip

\begin{figure}[htp]
\includegraphics[width=4.5in,clip]{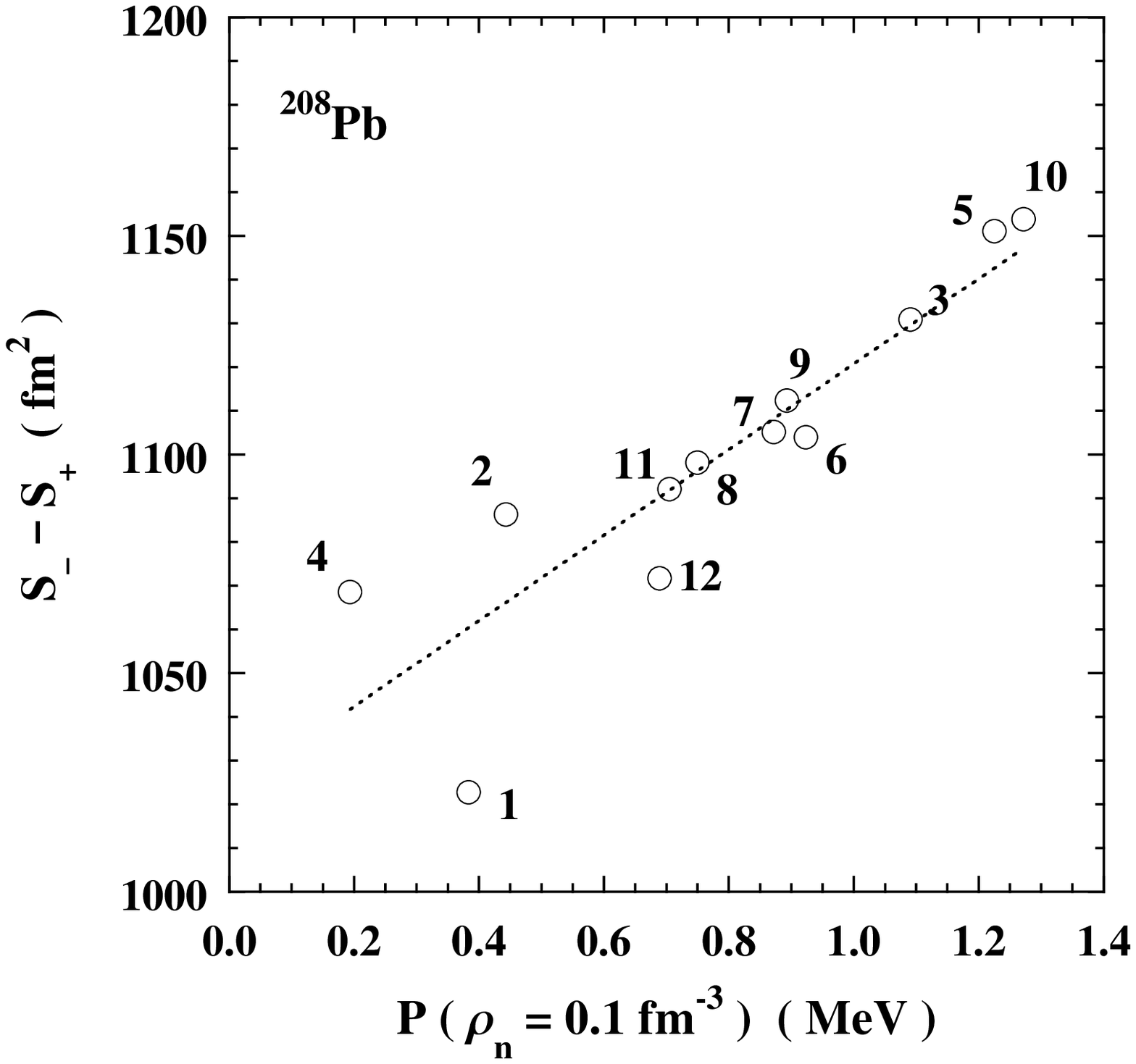}
\caption{\label{fig:pb208_np}
Correlations between the pressure of neutron matter and the SD sum rule 
values of $^{208}$Pb with 12 different Skyrme interactions. The numbers
 denote different Skyrme parameter sets: 1 for SI, 2 for SIII, 3 for 
SIV, 4 for SVI, 5 for Skya, 6 for SkM, 7 for SkM$^{*}$, 8 for SLy4, 9 for MSkA, 
10 for SkI3, 11 for SkX and 12 for SGII. The dashed 
line represents the result obtained by the least-squares method. The correlation coefficient
 is found to be r = 0.888. 
}
 \end{figure}

The correlations between the pressure of 
 neutron matter at the neutron density $\rho_n$ = 0.1 fm$^{-3}$
  and the charge exchange 
 SD sum rules of $^{90}$Zr and $^{208}$Pb 
 are shown in Figs. \ref{fig:zr90_np} and \ref{fig:pb208_np} 
with 12 different Skyrme interactions. The numbers
 denote different Skyrme parameter sets: 1 for SI, 2 for SIII, 3 for 
SIV, 4 for SVI, 5 for Skya, 6 for SkM, 7 for SkM$^{*}$, 8 for SLy4,
 9 for MSkA, 
10 for SkI3, 11 for SkX and 12 for SGII.  
The correlation coefficients from the extrapolated lines are
 r = 0.888  and 0.811 
for $^{208}$Pb and $^{90}$Zr, respectively. The correlation coefficients are
 somewhat smaller than those of the calculated correlation between
 the neutron skin thickness   
 $\delta_{np}$ and the pressure $P$ in ref. \cite{Yoshi04},
 but still, we can see fairly good correlations in Figs. \ref{fig:zr90_np}
 and \ref{fig:pb208_np}. 

The rms proton, charge, and neutron radii in $^{90}$Zr 
calculated by the HF model with the four interactions
SIII, SGII,  SkI3 and SLy4 are shown in Table \ref{tab:hf-zr}. 
The calculated charge radii of the SGII and SkI3 interactions 
 show reasonable agreement  with the experimental values.  However,  there 
 is a factor 2 difference in the neutron skin thickness $\delta_{np}$ 
 between the two interactions. 
 As seen in Table \ref{tab:hf-zr}, 
the neutron skin thickness $\delta_{np}$ obtained by the SD sum rules
 is consistent with the value previously obtained from
 the proton scattering data. However, the experimental uncertainty
 in the value 
 $\delta_{np} = (0.07\pm0.04)$ fm obtained by the SD sum rules is 
half that obtained through the proton data. This small uncertainty will 
help to disentangle the neutron matter EOS using the strong 
correlation with the neutron skin thickness.  
The experimental skin thickness 
 $\delta_{np} = 0.07\pm0.04$fm is close to the HF results
  of SLy4, as well as SGII and SIII.
The SkI3 result is not favored over the empirical result, even taking  the experimental uncertainties into 
consideration. We should also note that the experimental peak energy 
of $t_-$ SD strength 
  in $^{90}$Nb coincides with the calculated peak energy of the SLy4 
 interaction,  while that of SkI3 is 4 MeV above the experimental value, 
  as seen in Fig. \ref{fig:zr90_sd_exp}.
While all interactions lie within the experimental value 
 $\Delta S=(147\pm 13)$fm$^2$ in Fig. \ref{fig:zr90_np},
 the empirical data favor the interactions indicated by the numbers  2(SIII),
 11(SkX), 8(SLy4), 7(SkM*) and 6(SkM).
 These interactions suggest a soft neutron matter EOS
  with the pressure P($\rho_n$=0.1fm$^{-3}$) = (0.65$\pm$0.2) MeV.  
 Thus, the preferred nuclear matter symmetry energy extracted 
 from the SD experiment 
  is found to be 
 J = (30$\pm$2) MeV as a result of the strong correlation between the
 neutron skin thickness and the symmetry energy \cite{Furn,Dani}.

Table  \ref{tab:hf-pb} tabulates the rms proton, charge 
 and neutron radii in $^{208}$Pb  
calculated by the 
  HF model, along with the experimental charge radius. 
The HF results of SGII and SLy4 account for
the experimental charge radius, while there is a large variation in the predictions for the neutron skin thickness $\delta_{np}$. 
The empirical value of the neutron skin thickness 
$\delta_{np}$ in $^{208}$Pb was 
obtained by proton scattering experiments.
However, the values obtained depend very much on the experiments
 and analyses.  That is, 
the experimental errors are still large and some of 
the values obtained have 
no overlap, even when the uncertainty in the analyses is taken into account;  
  $\delta_{np} = (0.14 \pm 0.02)$ fm in ref. 
\cite{Hoffman}, $\delta_{np} = (0.20 \pm 0.04)$ fm  in
   ref. \cite{Starodubsky}
 and $(0.083 < \delta_{np} < 0.111)$ fm  in ref. \cite{pscatt2}.
 We quote in Table \ref{tab:hf-pb} the  value in ref. \cite{pscatt2}
 where the analyses were
  performed comprehensively with many different sets 
 of data including those adopted in refs. \cite{Hoffman,Starodubsky}.
Although these results depend on the effective nucleon-nucleon
effective interactions in nuclei used in the analysis, 
the  comprehensive study of proton scattering
  in ref. \cite{pscatt2}  
 reports rather small neutron skin thicknesses, even smaller than the
 smallest value in Table 
 \ref{tab:hf-pb} obtained using the SIII interaction.  Again, this small  $\delta_{np}$ 
 suggests a   
 soft neutron matter EOS similar to the conclusion reached by the
SD sum rules of $^{90}$Zr. 
The charge exchange $^{208}$Pb($^3$He,$t$)$^{208}$Bi 
 reaction data \cite{Aki}
show an SD peak  in $^{208}$Bi
 at $E_{x}$ = 24.8$\pm$0.8 MeV measured 
from the ground state of $^{208}$Pb, as marked by an arrow
 in Fig. \ref{fig:pb_sdmp}.  This peak position is close to the 
calculated value of the SGII interaction, while the SkI3 peak is a few 
MeV higher than the empirical value. This comparison may exclude the prediction
by SkI3, which gives a hard neutron matter EOS in Fig. \ref{fig:pb208_np} 
marked by the number 10.

 The neutron skin thickness was determined by the giant dipole 
resonance experiment to be $\delta_{np}=(0.19 \pm0.09)$fm \cite{GDR}.   
This analysis depends on 
the adopted transition density and also 
the optical potentials so that the result is highly
model-dependent. We  
 definitely need more quantitative information, i.e., 
    model-independent information on the
 neutron skin thickness in $^{208}$Pb 
 for precise determination of the neutron matter EOS as well as  the 
 isovector nuclear matter properties.  To this end, the charge
 exchange SD experiments of $^{208}$Pb will provide useful model-independent 
 information with the same accuracy as the parity violation electron 
 scattering experiment.

\section{Summary} 
We have investigated the  SD excitations in $^{90}$Zr and $^{208}$Pb
using the    
HF + RPA model    with four 
Skyrme interactions, viz., SIII, SGII, SkI3 and SLy4.  
  It is shown that 
the Landau damping effect plays an important role in explaining
  the large observed width of SD resonance, while
 the coupling to the continuum is rather weak. 
     Among the four interactions, the peak position of the
 experimental $t_-$  SD strength in $^{90}$Nb 
 is well described by the SLy4 interaction, 
while the results of SIII and SGII are  also acceptable.
For the  $t_+$ excitation of   $^{90}$Zr, 
a two-peak structure was found in both 
the experimental and calculated results.  The  
 SLy4 and SkI3 results showed good 
agreement with  the observed low energy peak.
We pointed out that the calculated results need a quenching factor 
 quf$\simeq$0.68 to allow a quantitative comparison with the 
experimental data up to $E_x$ = 36(40) MeV for the $t_- (t_+)$ channel in Fig. 
\ref{fig:zr90_sd_exp}.  
About 30\% of the NEWSR value is found in the excitation energy above 
$E_x = $36(40) MeV for the $^{90}$Zr(p,n) $^{90}$Nb ($^{90}$Zr(n,p) $^{90}$Y)
   experiments.  
  The calculated SD sum rule $\Delta S=S_- -S_+$
shows good saturation properties above $E_x = $40 MeV without any 
 quenching factor relative to
 the observed data despite the fact that sum rules $S_-$ and $S_+$ 
  themselves  
 increase gradually above $E_x\geq$40 MeV.
 The neutron skin  thickness $\delta_{np} = 0.07\pm0.04$ fm 
  extracted from the SD sum rules is 
close to the calculated values obtained using SLy4 as well as SIII and SGII.
However, the extracted value does not favor 
the  SkI3 interaction which gives almost twice as large a neutron
skin thickness as SIII and SGII. 
 This is indicative of the soft neutron matter EOS induced by the 
strong linear correlation between the neutron matter EOS and the neutron
skin thickness.
 We showed that the SD strength of the $t_-$ excitation of $^{208}$Pb 
 has a large width due to the Landau damping effect.
In contrast, the $t_+$ excitation of $^{208}$Pb turns out to be 
a single peak in a rather low energy region because of the strong
Pauli blocking effect of the excess neutrons.
  The  peak of the $t_-$ SD strength was observed by $^{208}$Pb($^3$He,t)
 $^{208}$Bi at $E_x\sim$25 MeV.  This peak energy
  coincides with the peak calculated using 
the SGII interaction, while the SkI3 interaction yields a peak that is a few MeV higher 
than the empirical peak.
 Thus,
 the empirical SD sum rule values of $^{90}$Zr and the observed peak 
energies of the $t_-$ SD strength distributions in $^{90}$Nb and $^{208}$Bi
 indicate a soft neutron matter EOS with 
a pressure of 
 P($\rho_n$=0.1fm$^{-3})$ = (0.65$\pm$0.2) MeV.  
 The  nuclear matter symmetry energy is also 
  determined to be 
 J = (30$\pm$2) MeV from the strong correlation between the
 neutron skin thickness and the symmetry energy. 
In order to draw a more definite conclusion on the SD sum rules, 
 as well as the neutron skin thickness and the neutron matter EOS, 
 we need quantitative experimental work to obtain    
the SD sum rules in heavy nuclei like $^{208}$Pb, both in 
the $t_-$ and $t_+$ channels.  

\section*{Acknowledgments}
The authors would like express their thanks to I. Hamamoto  
for useful discussions. 
This work was supported in part by  Grant-in-Aid for Scientific Research 
No. 16540259 and No. 17002003 from 
the Ministry of Education, Science, Culture and Sports and
 National Natural Science Foundation of China (No.10605018). 







\end{document}